\documentclass[12pt]{interact}

\usepackage{epstopdf}
\usepackage[caption=false]{subfig}
\usepackage[nolists,tablesfirst]{endfloat}

\DeclareDelayedFloatFlavor{sidewaysfigure}{figure}
\DeclareDelayedFloatFlavor{sidewaystable}{table}

\usepackage[british]{babel}
\usepackage[title]{appendix}
\usepackage{amsmath}
\usepackage{amssymb,amsfonts,bm}
\usepackage[section]{placeins}
\usepackage{array}
\usepackage{lipsum,color,float}
\usepackage{footnote}
\usepackage[usenames,dvipsnames,svgnames,table]{xcolor}
\usepackage{hyperref}
\usepackage{enumerate}
\usepackage{graphicx}
\usepackage{epstopdf}
\usepackage{graphics}
\usepackage{subfiles}
\usepackage{listings}

\usepackage[numbers,sort&compress]{natbib}
\bibpunct[, ]{[}{]}{,}{n}{,}{,}
\renewcommand\bibfont{\fontsize{10}{12}\selectfont}
\makeatletter
\def\NAT@def@citea{\def\@citea{\NAT@separator}}
\makeatother

\theoremstyle{plain}
\newtheorem{theorem}{Theorem}[section]

\theoremstyle{definition}

\theoremstyle{remark}

\newcommand{\m}{\mbox}

\newcommand{\trn}{\operatorname{trn}}

\newcommand{\sgn}{\operatorname{sgn}}
\newcommand{\hyp}{\operatorname{hyp}}
\newcommand{\lch}{\operatorname{lch}}
\newcommand{\lne}{\operatorname{lne}}
\newcommand{\R}{\mathbb{R}}

\begin{document}


\title{A family of smooth piecewise-linear models with probabilistic interpretations}

\author{
\name{I.E.P. Ferreira\textsuperscript{a}\thanks{CONTACT I.E.P. Ferreira. Email: iuri@ufscar.br} and S.S. Zocchi\textsuperscript{b}}
\affil{\textsuperscript{a}Centro de Ci\^encias da Natureza, Universidade Federal de S\~ao Carlos, Buri-SP, Brazil; \textsuperscript{b}Departamento de Ci\^encias Exatas, Escola Superior de Agricultura "Luiz de Queiroz", Piracicaba-SP, Brazil.}
}

\maketitle

\begin{abstract}
The smooth piecewise-linear models cover a wide range of applications nowadays. Basically, there are two classes of them: models are transitional or hyperbolic according to their behaviour at the phase-transition zones. This study explored three different approaches to build smooth piecewise-linear models, and we analysed their inter-relationships by a unifying modelling framework. We conceived the  smoothed phase-transition zones as domains where a mixture process takes place, which ensured probabilistic interpretations for both hyperbolic and transitional models in the light of random thresholds. Many popular models found in the literature are special cases of our methodology. Furthermore, this study introduces novel regression models as alternatives, such as the Epanechnikov, Normal and Skewed-Normal Bent-Cables.
\end{abstract}

\section{Introduction}
\noindent 
Piecewise-linear models are used in many areas of knowledge, such as 
medicine \cite{Muggeo2003,Segalas2020}, agriculture \cite{Cerrato1990,Prunty1983,Berck1990,ferreira2017},  ecology \cite{Toms2003,LONGATO2018}, environmental sciences and economy \cite{Khan2018}, and physics \cite{Jimenez-Fernandez2016a, Roslyakova2017}. 
The original, abrupt piecewise-linear model is expressed by  
\begin{eqnarray}
\label{modelo1}
y_i = \eta(x_i) + \epsilon_i ~~  (i=1,2,\cdots,n), \nonumber
\end{eqnarray}
where the fixed part of the model, $\eta$, is sliced into 
$D$ intercepting straight lines, also called phases or 
regimes, according to the values of the explanatory 
variable $x$. Then, the phases are expressed by 
$g_j(x) = \alpha_j + \beta_j \, x$, for $j=1, \cdots, D$ 
and $\tau_{j-1} \leq x \leq \tau_{j}$, where $\tau_j$'s 
are the change-points (thresholds). The $\alpha$'s, $\beta$'s 
and $\tau_1, \tau_2, \cdots, \tau_{D-1}$ are free parameters 
to be estimated, while the first and the last change-points can 
be conveniently fixed according to the $x$ range, that is 
$\tau_0 = x_{(1)}$ and $\tau_{D} = x_{(n)}$.

The continuity of the response curve is ensured by taking
$g_l(\tau_l) = g_{l+1}(\tau_l)$, for $l=1, \cdots, D-1$. However, 
the joining of straight lines causes discontinuities on first-order 
derivatives of the model at the change-points. Furthermore, the change-points 
often are unknown, and it characterises a complicated type of nonlinear 
model for which standard procedures of estimation and inference 
do not hold. Optimisation problems become simpler when the abrupt model is replaced by a smooth approximating function \cite{SW2003}. Indeed, it can be considered a more realistic approach when data exhibit some degree of smoothness at transition, instead of a sudden change.  

One of the first smooth approximations for piecewise-linear models 
was introduced by Bacon and Watts \cite{Bacon1971}, in an attempt to overcome 
non-differentiability and provide a more reliable test for the 
existence of structural change (i.e. test two phases 
against only one). They wrote the piecewise model using a sign function 
to separate the phases; then, they replaced the sign by smooth sigmoid 
functions, such as the hyperbolic tangent.
In the years that followed, many similar alternatives to smoothing the piecewise-linear 
models appeared in the literature. The idea is always the same  - authors generally suggest families of smooth approximating functions in order to replace the abrupt operators (maximum, modulus, or sign) in the original (abrupt) piecewise models \cite{Bacon1971,Griffiths1973,Watts1974,Griffiths1975,TZ1981a,TZ1981b,TZ1983,SW2003,Lazaro2001,chiu2006,Jimenez-Fernandez2016a,Khan2018}. In this way, the linear phases are joined by a curved transition, with an adjustable degree of smoothness (radius of curvature). 
The smoothing strategies were proved to be useful in many ways, 
such as to avoid the lack of fit, measure the degree of 
smoothness from data, and even to specify the 
limits of the phase-transition zone \cite{Toms2003,Khan2018}. 

Basically, there are two classes of smooth piecewise-linear models, the hyperbolic and transitional models \cite{SW2003}. This classification depends on the graph's shape in the phase-transition zone: transitional models show an unnatural `bulge' formation, while hyperbolic models show a gradually inflected transition \cite{Griffiths1975}. Bacon and Watts's hyperbolic tangent model is a famous example of the transitional model \cite{Bacon1971}, while the bent-cable regression of Chiu et al. is the most popular hyperbolic model \cite{chiu2006}.

This study explored three different approaches for smooth piecewise-linear 
modelling. These approaches are inter-related and allow us to assign 
probabilistic interpretations for both the hyperbolic and transitional 
models. The proposed methodology is a unifying framework for piecewise-linear 
modelling based on the assumption of phase-transition random thresholds. From 
this perspective, the smooth transition zones are domains where a mixture of 
processes takes place, and the degree of smoothness required in the model 
is simply a consequence of that. 

Many models found in the literature were 
proven to be special cases of our methodology when considering 
specific distributions for the thresholds. As examples, they can be mentioned the 
bent-cable model linked with the Uniform distribution, the Griffiths and Miller's 
model linked with the non-standard t-distribution, and the Bacon and Watts's model 
linked with the Logistic distribution. We also provided new bent-cable models 
based on the Epanechnikov, Normal and Skew-Normal distributions.

The text is organised as follows: section \ref{syn} brings a brief review of the theoretical foundations for piecewise-linear modelling. The methodological section explores three different ways to derive these models (section \ref{methodology}): 
\begin{itemize}
    \item The first approach consists of an extension of the Chiu's bent-cable regression that allows the use of higher degree polynomials as bent-functions. We called it the Extended Bent-Cable (Ext. BC) approach (section \ref{SEC-EXT});
\item The second method is a State Mixture Modelling (SMM) approach, based on the assumption of random thresholds. The equivalence between the SMM approach and transitional models is stated in section \ref{MM};
\item The third method may be viewed as a correction for transitional models, that turns them into their hyperbolic-shaped version. In other words, all the transitional models can be corrected by an adding factor, becoming a hyperbolic model. This type of hyperbolic models will be called the Expected Bending-Cable (Exp. BC) model hereinafter (section \ref{EBCM}).
\end{itemize}
Section \ref{SEC-INTER} introduces new bent-cable models and discussions about the model inter-relationships. Section \ref{estimation} presents the standard procedures for classical estimation and inference. Section \ref{application} presents a comparative analysis of bent-cables models using R. A. Cook's data on the stagnant surface layer height as a practical motivation. All technical proofs are placed in the appendices.
\section{\label{syn} Background review and synthesis of the literature}
Basically, there are two popular ways to express mathematically the 
piecewise-linear models - the sign and max-min formulations. A brief 
description of these techniques is presented below. 
 
\subsection{The sign formulation - Transitional and hyperbolic models}
According to Bacon and Watts \cite{Bacon1971}, two adjacent regimes can be joined 
as follows
\begin{eqnarray}
\label{sinal}
\eta(x)  &=& \theta_0 + \theta_1 (x-\tau) + \theta_2 |x-\tau| \nonumber \\
       &=& \theta_0 + \theta_1 (x-\tau) + \theta_2 (x-\tau) \sgn(x-\tau),
\end{eqnarray}
where $\tau$ is the change-point, $\theta_0=\alpha_1 + \beta_1 \tau$, $\theta_1 = (\beta_1+\beta_2)/2$, and $\theta_2=(\beta_2-\beta_1)/2$ (Fig. \ref{fig1correcao}). Note that the sign ($\sgn$) of 
$\theta_2$ determines if the curve will be convex ($\theta_2>0$) or concave ($\theta_2<0$). 

\setcounter{figure}{1}
\begin{sidewaysfigure}[p]
\begin{minipage}{11cm}
\includegraphics[width=10cm,height=10cm]{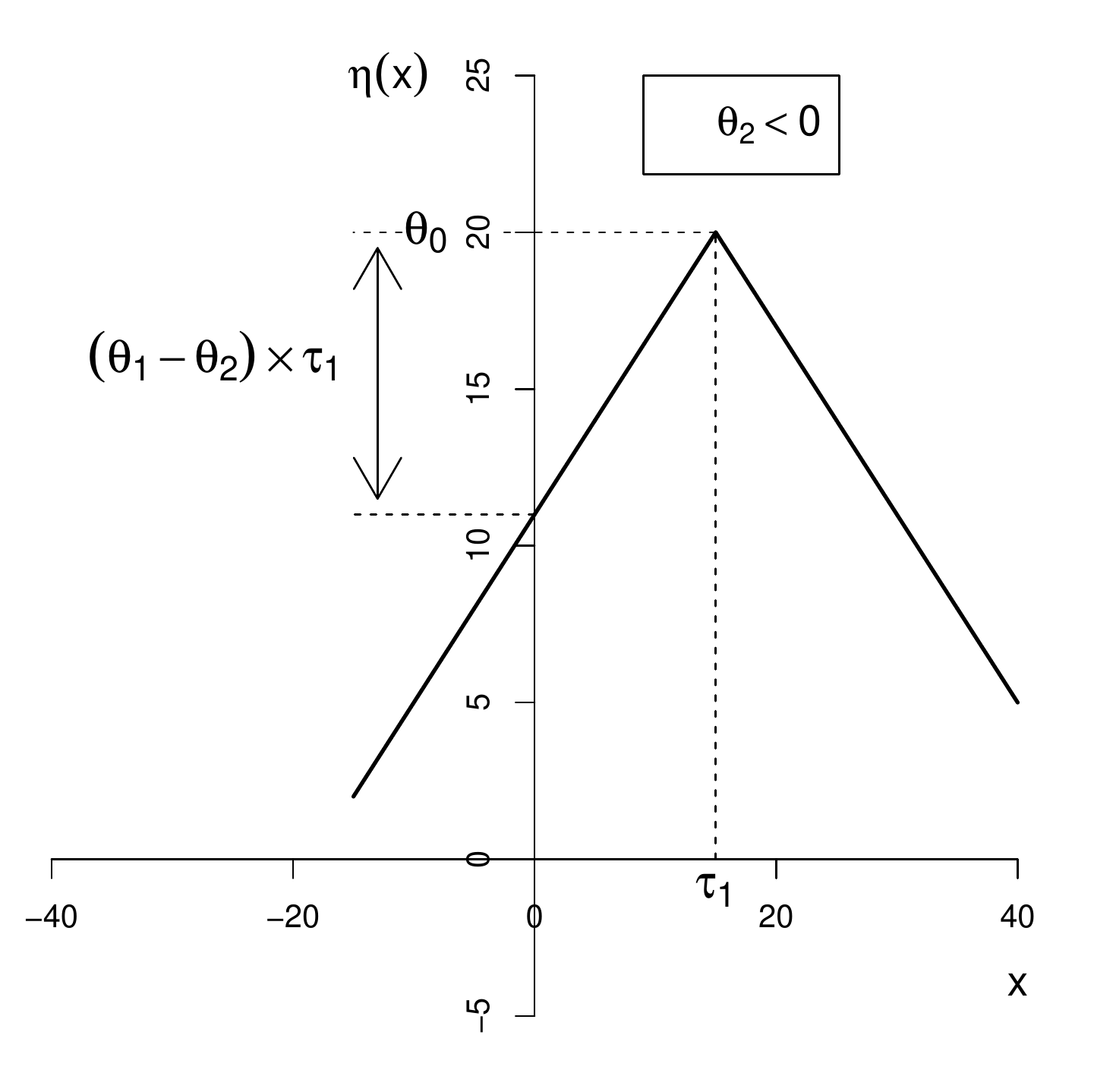}
\end{minipage}
\hfill
\begin{minipage}{11cm}
\includegraphics[width=10cm,height=10cm]{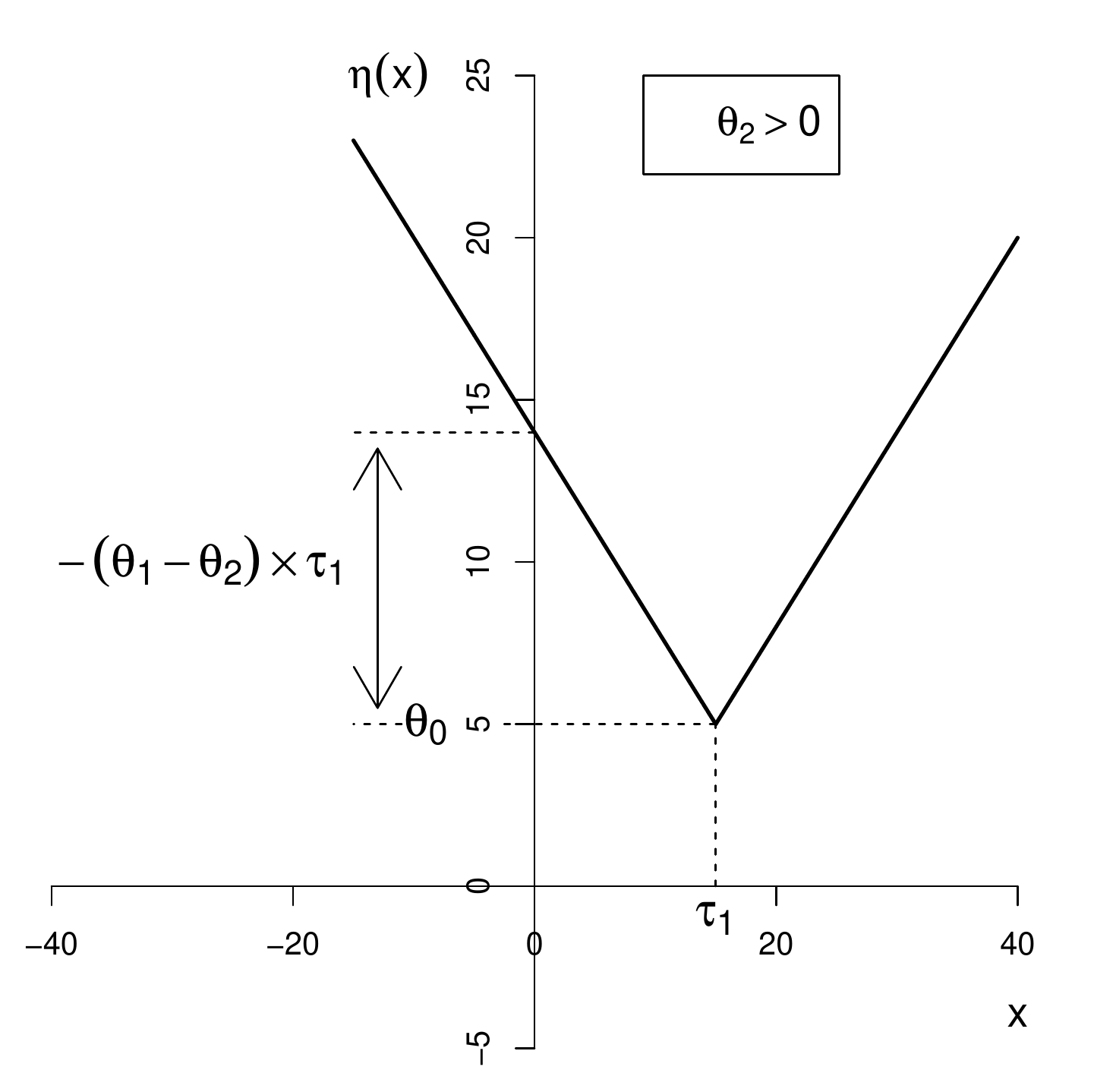}
\end{minipage}
\caption{\label{fig1correcao}Two straight lines joined by `sign formulation' considering a convex ($\theta 2>0$) and concave ($\theta 2<0$) response.}
\end{sidewaysfigure}

This formulation employs the sign (or modulus) function to connect the 
adjacent linear phases and, for this reason, it became known as the `sign formulation' \cite{SW2003}. Based on that, two families of smoothing approximations were derived: 
the Bacon and Watts's transitional models and Griffiths and Miller's hyperbolic models \cite{Bacon1971,Griffiths1973,Watts1974,Griffiths1975}.  

The piecewise model with an abrupt transition (Eq. \ref{sinal}) forms two conjugate angles (adding up 360º) at the intersection of the straight lines. The transitional and 
hyperbolic models differentiate from each other in the way the response curve  
bends under this region (Fig. \ref{hypandtrans}). 

\setcounter{figure}{1}
\begin{sidewaysfigure}[p]
\begin{minipage}{11cm}
\includegraphics[width=10cm,height=10cm]{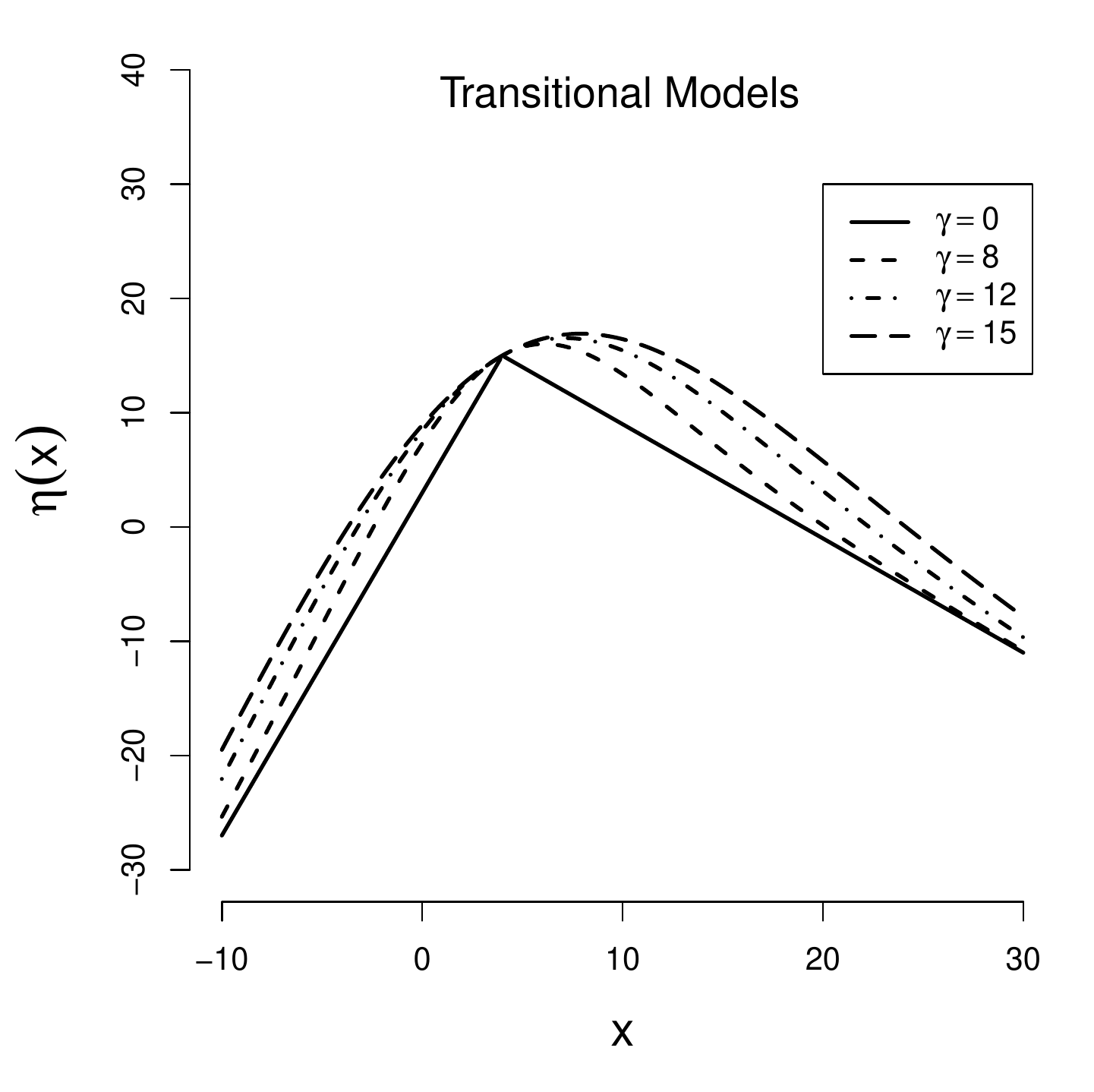}
\end{minipage}
\hfill
\begin{minipage}{11cm}
\includegraphics[width=10cm,height=10cm]{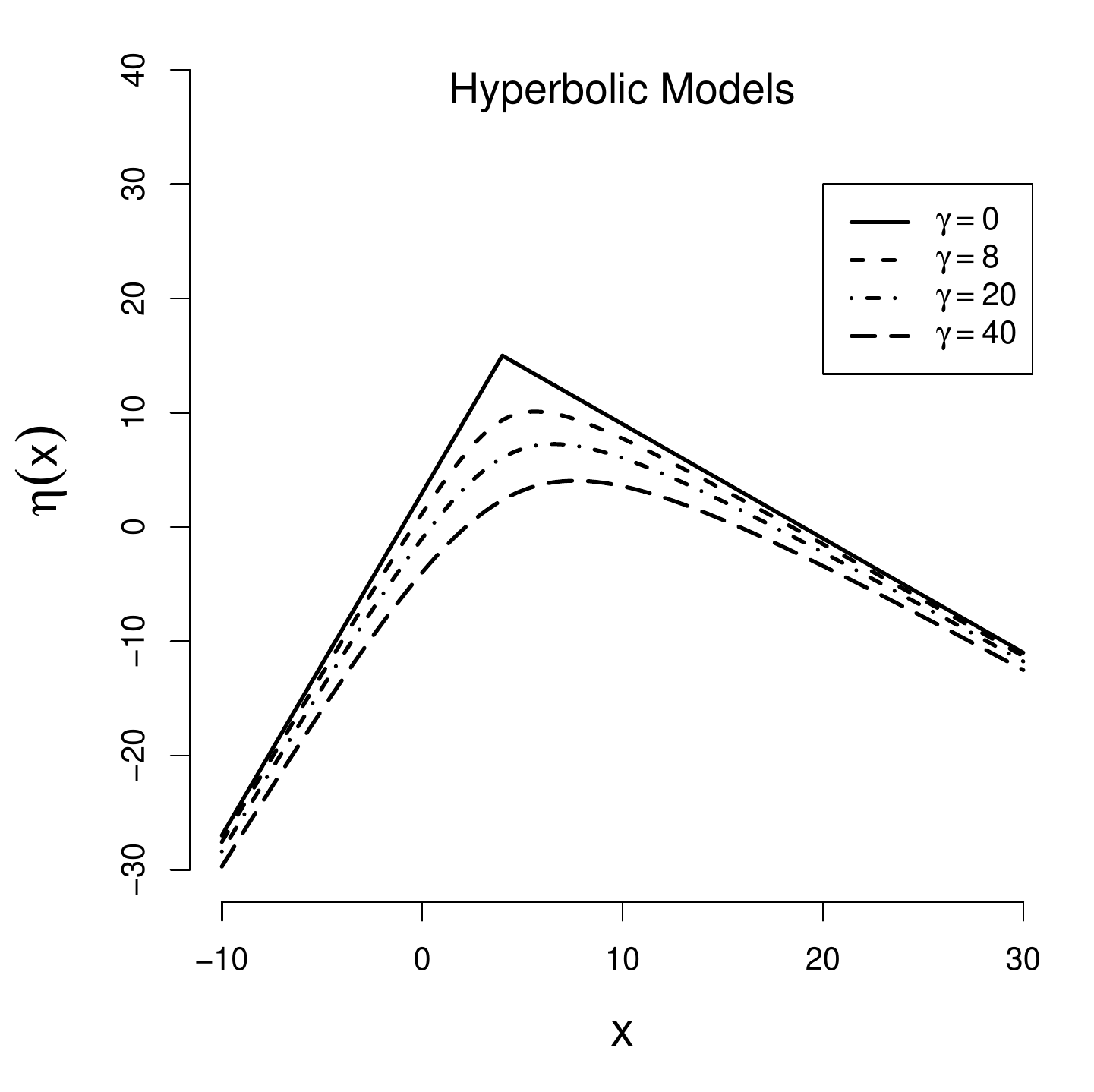}
\end{minipage}
\caption{\label{hypandtrans}
The graphs of the original transitional and hyperbolic models \cite{Bacon1971,Griffiths1973} for different $\gamma$ values.}
\end{sidewaysfigure}

The graph of transitional models crosses the greater angle, and it always touches the intersection point, what leads to the `bulging phenomenon' - that is, the response curve shows an odd protuberance (or `bulge') near the transition point. On the other hand, the graph of hyperbolic models crosses the smaller angle, and it does not touch the intersection point - response curve approaches to the straight lines just as the hyperbola approaches their asymptotes. Despite this difference, the transitional and hyperbolic models have the same asymptotic behaviour when $x$ moves 
away from $\tau$. For both the models, there is a smoothing parameter 
($\gamma > 0$) that controls the radius of curvature at the 
intersection point. The abrupt model could be considered as a 
particular case when $\gamma \rightarrow 0^{+}$. 

In their seminal work, Bacon and Watts \cite{Bacon1971} proposed a 
family of smooth transitional functions to replace the sign 
function ($\sgn$) in Eq. (\ref{sinal}). According to the literature \cite{Griffiths1973,SW2003}, any smooth function ($C^1$ class)
satisfying the following three conditions is transitional:    
\begin{itemize}
\item[(c.1)]{$\displaystyle{\lim_{s \rightarrow \pm \infty}}  \, [ s \, \trn(s) - |s|] = 0$,}
\item[(c.2)]{$\displaystyle{\lim_{\gamma \rightarrow 0^{+}}} \trn(s / \gamma)= \sgn(s)$, }
\item[(c.3)]{$\trn(0) = \sgn(0) = 0$.}
\end{itemize}
The condition (c.1) ensures the appropriate asymptotic behaviour when $s$ moves away from the origin. The condition (c.2) ensures that the abrupt model is a special case when $\gamma \rightarrow 0^{+}$. By the last, the condition (c.3) 
ensures that the smooth model passes through the intersection point $(\tau, \theta_0)$. 

Note that the condition (c.3) is too restrictive, and to some extent, it is unnecessary. The argument $s$ of the transitional function reaches zero at the intersection point, with $x=\tau$. Then, all the terms that are multiplied by $\theta_1$ and $\theta_2$ in Eq. (\ref{sinal}) vanish, and any finite 
value for $\trn(0)$ gives the same result. 
Thus, without loss of generality, we slightly modified the conditions 
for transitional models: 
\begin{itemize}
\item[(i)]{$\displaystyle{\lim_{x \rightarrow -\infty}} (x-\tau) \, [\trn(x-\tau, \gamma) +1] = 0 ~~~ \mbox{and} ~~~ \displaystyle{\lim_{x \rightarrow +\infty}} (x-\tau) \, [\trn(x-\tau, \gamma) - 1] = 0$,}
\item[(ii)]{ $\displaystyle{\lim_{\gamma \rightarrow 0^{+}}} \trn(x-\tau,  \gamma)= \sgn(x-\tau)  ~~~~~  \mbox{for} ~~~ x \neq \tau$, }
\item[(iii)]{$|\trn(0, \gamma)| < +\infty$.}
\end{itemize}

Many functions satisfy these conditions, and the hyperbolic tangent is the most common choice:
\begin{eqnarray}
\trn(x-\tau,\gamma)=\tanh[(x-\tau)/\gamma]. \nonumber
\end{eqnarray}
Polynomials can also be used to build transitional functions. Based on the Lin and Unbehauen's study \cite{Lin1993}, L\'azaro et al. \cite{Lazaro2001} adopted a 
6$^{th}$-degree polynomial to provide a smooth approximation for the modulus 
function, $|x-\tau|$. The corresponding smooth approximation for the $\sgn(x-\tau)$ 
is giving when it is divided by $(x-\tau)$, resulting in 
\begin{eqnarray}
\label{sextaordem}
\trn(x-\tau, \gamma) = \left \{
\begin{array}{lr}
-1 & x - \tau < - \gamma, \\
\frac{1}{8\, \gamma} \Big [ 15 \, (x-\tau ) - 
10 \,  \frac{(x-\tau)^3}{\gamma^2}   + 3\, \frac{(x-\tau)}{\gamma^4}^5 \Big ]  ~~ &  |x-\tau| \leq \gamma, \\
1 & x -\tau > \gamma.
\end{array}
\right . 
\end{eqnarray} 
This approximation was previously shown by Bunke and Schulze \cite{Bunke1985}, based on the Tishler and Zang's ideas that will be the main subject in the next section \cite{TZ1981a}.    

Alternatively, Bacon and Watts also suggested the use of the cumulative 
distribution function (CDF) of any symmetric probability 
density function (PDF) \cite{Bacon1971}. Obviously, they were 
considering taking 
\begin{eqnarray}
\trn(x-\tau, \gamma)= 2\,F(x; \tau, \gamma) - 1 , \nonumber
\end{eqnarray}
where $F$ is the CDF. In fact, there are few conditions that need to be met for a 
CDF to be a transitional function, and 
some distributions are unsuitable for this task. 
Indeed, it is possible to derive transitional approximations 
based on asymmetrical distributions, contrary to what Bacon and Watts said. 
We will address these issues in the methodology (section \ref{methodology}).   

The transitional models show a distortion (the `bulge') in the phase-transition zone; for this reason, they were criticised by many authors\cite{Griffiths1973,Watts1974,SW2003,Jimenez-Fernandez2016a}. Often the data sets do not show any evidence for this `bulge' aspect in nature, and this distortion may be considered an aberration arising from the model. 

Griffiths and Miller \cite{Griffiths1973} suppressed the third condition; 
in this way, they provide an approximation that does not show this `bulging phenomenon'. They proposed the following bent-function:   
\begin{eqnarray}
\hyp(x-\tau, \gamma) =  \frac{\sqrt[]{(x-\tau)^2+\gamma}}{x-\tau}.  \nonumber
\end{eqnarray}
This function has a discontinuity at $x=\tau$ that is solved when the term 
replaces the $\sgn$ in the model (Eq. \ref{sinal}). This approximation satisfies 
condition (i), ensuring its asymptotic behaviour. Furthermore, the condition (ii) is met 
since $\gamma \rightarrow 0^{+}$ implies $\sqrt[]{(x-\tau)^2+\gamma} \rightarrow |x-\tau|$ and, consequently, $\gamma \rightarrow 0^{+}$ leads to the original abrupt model 
in Eq. (\ref{sinal}). However, the value of the hyperbola approximation diverges to infinity as $x$ approaches to $\tau$, and the third condition [(c.3) or (iii)] 
does not hold in this situation. Models built on this approximation are called hyperbolic models \citep{SW2003}. 

The graphs of transitional and hyperbolic approximations have notable 
differences, which explains the discrepancies in the smooth model shapes (Fig. \ref{hypandtrans}). The transitional shape only arises if 
$(x-\tau) \, \trn(x-\tau, \gamma) \leq |x-\tau|$ or, equivalently, 
$\trn(x-\tau, \gamma) \geq -1$ for $x \leq \tau$ and 
$\trn(x-\tau, \gamma) \leq 1$ for $x\geq \tau$. For this 
reason, the common transitional functions found in the
literature assume values belonging to $[-1,1]$. Indeed, the 
useful transitional functions are monotonic 
(non-decreasing), otherwise undesirable fluctuations would 
be seen in the model, under the transition zone. On the other hand, 
the hyperbolic shape only arises if $(x-\tau) \, \hyp(x-\tau, \gamma) \geq |x-\tau|$, 
that is $\hyp(x-\tau, \gamma) \leq -1$ for $x \leq \tau$ 
and $\hyp(x-\tau, \gamma) \geq 1$ for $x\geq \tau$. For this 
reason, the hyperbola-like approximation assumes values 
belonging to $(-\infty, -1]U[+1,+\infty)$. The hyperbolic approximations 
also differ from the transitional ones for having a non-monotonic behaviour, in 
addition to its removable discontinuity at $x=\tau$ (Fig. \ref{fig2}).   

\setcounter{figure}{1}
\begin{sidewaysfigure}[p]
\begin{minipage}{11cm}
\includegraphics[width=10cm,height=10cm]{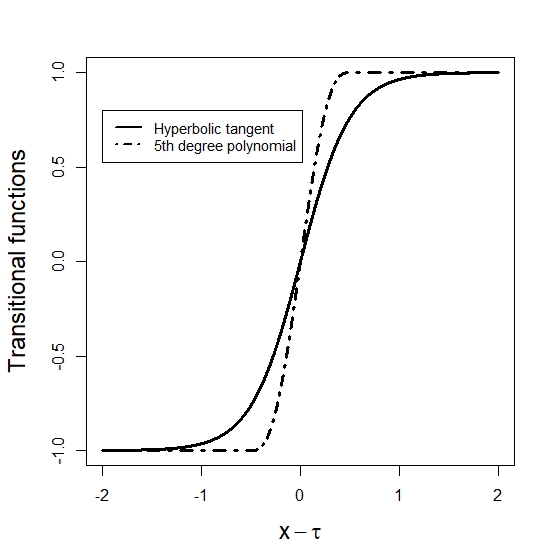}
\end{minipage}
\hfill
\begin{minipage}{11cm}
\includegraphics[width=10cm,height=10cm]{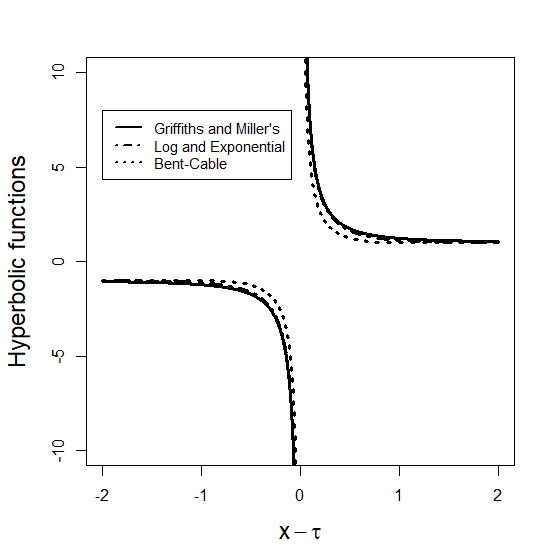}
\end{minipage}
\caption{\label{fig2} The most common transitional (at the left) and hyperbolic (at the right) 
functions. Transitional: the hyperbolic tangent \cite{Bacon1971}, and the 5$^{th}$ degree polynomial functions \cite{Bunke1985,Lazaro2001}. Hyperbolic: the hyperbolic model \cite{Griffiths1973}, the log and exponential \cite{Lazaro2001}, and the original bent-cable model \cite{SW2003}. The smoothing parameters were set as $\gamma=0.5$, and $\zeta=1.0$. }
\end{sidewaysfigure}

Considering this characteristic behaviour, other hyperbola-like 
approximations could be provided. In this study, the hyperbola-like 
approximations ($\hyp$) are defined as follows: 
\begin{itemize}
    \item[(h.1)]{The $\hyp(x-\tau,\gamma)$ satisfies the conditions (i) and (ii),}
    \item[(h.2)]{The $\hyp(x-\tau,\gamma)$ does not satisfies (iii),}
    \item[(h.3)]{$|\hyp(x-\tau, \gamma)| \geq 1$,}
    \item[(h.4)]{$(x-\tau) \, \hyp(x-\tau, \gamma)$ is differentiable over $x \in \R$.}
\end{itemize}
Models based on functions that satisfy all these conditions will be called 'hyperbolic models' hereinafter. Many models used in these days fall into that category.  

For example, Jimenez-Fernandez et al. \cite{Jimenez-Fernandez2016a} suggested the Log and Exponential approximation ($\lne$) for the modulus:
\begin{eqnarray}
|x-\tau| \approx \lne(x-\tau, \gamma) &=&  2 \, \gamma \log(e^{-(x-\tau)/2\, \gamma}+ e^{(x-\tau)/2\, \gamma}) \nonumber \\ 
&=& 2 \, \gamma \, \log(e^{(x-\tau)/\gamma}+1) - x +\tau. \nonumber
\end{eqnarray}
This approximation has the advantage to provide continuous n$^{th}$-order derivatives, 
and simplifications are easier since it is based on elementary inverse functions - 
the exponential and the logarithmic functions. The $\lne(.)$ approximation is 
naturally linked with a hyperbolic model since
$\hyp(x-\tau, \gamma) = \lne(x-\tau,\gamma)/(x-\tau)$
satisfies the conditions (h.1-4). 

L\'azaro et al. \cite{Lazaro2001} introduced a similar approximation for the modulus, the 
Log Hyperbolic Cosine ($\lch$): 
\begin{eqnarray}
\lch(x-\tau, \gamma) = \gamma \, \log\{ \cosh[(x-\tau)/\gamma]\}.  \nonumber
\end{eqnarray}
The $\lch(.)$ cannot be considered as a hyperbolic transition, as we stated, 
since $\lch(x-\tau, \gamma) / (x-\tau)$ approaches zero when 
$x \rightarrow \tau$. Also, it cannot be considered a transitional 
model since  $\lch(x-\tau, \gamma) / (x-\tau)$ does not
satisfy the condition (i) for $\trn(.)$ functions: 
\begin{eqnarray}
\displaystyle{\lim_{x \rightarrow  \pm \infty}} \lch(x-\tau, \gamma) - |x-\tau| = -\gamma \, \log(2),
~~~~ \mbox{for} ~~\gamma>0. \nonumber 
\end{eqnarray}
It is easy to show that the $\lch$ function is a vertical translation of $\lne$:  
\begin{eqnarray}
 \lch(x-\tau, \gamma^*) &=& \gamma^* \, \log \Big [ \frac{1}{2} \, \Big ( e^{-(x-\tau)/ \gamma^*}+ e^{(x-\tau)/ \gamma^*} \Big ) \Big ] \nonumber \\
 &=&  \gamma^* \, \log \Big [ e^{-(x-\tau)/ \gamma^*}+ e^{(x-\tau)/ \gamma^*} \Big ]   
 - \gamma^* \log(2) \nonumber \\
 &=& \lne(x-\tau, \gamma^*/2) - \gamma^* \, log(2). \nonumber
\end{eqnarray}
Thus, models built on $\lch$ are $\lne$-hyperbolic models shifted  
until the curve touches the interception point. The displacement factor is 
$d = - \gamma^* \, log(2)$, and due to it, the modulus function is not well 
approximated - it causes a departure between the original linear regimes and 
the model asymptotes. Despite that, it still can be successfully used in practice \cite{Lazaro2001,Jimenez-Fernandez2016a}.
\subsection{The max-min formulation}
The max-min formulation is a flexible way to deal with problems in piecewise 
regression \cite{TZ1981b,TZ1981a}. This approach uses maximum and minimum 
operators to join adjacent phases, thus creating a continuous curve or 
surface. 

For example, the model for two-phase problems is given by
\begin{eqnarray}
\label{maxmin}
\eta(\mathbf{x},\m{\bm{$\beta$}}) = \max \{g_1(\mathbf{x}, \m{\bm{$\beta$}}_1), g_2(\mathbf{x}, \m{\bm{$\beta$}}_2) \}  ~~ \mbox{or} ~~ \min \{g_1(\mathbf{x}, \m{\bm{$\beta$}}_1), g_2(\mathbf{x}, \m{\bm{$\beta$}}_2) \}. \nonumber
\end{eqnarray}
The choice between the minimum and maximum depends on the concavity of the model. 
Three or more phases could be joined similarly, by using the maximum for convex 
regions and the minimum for concave ones.

There is no restriction concerning the sub-models' shape, and the phases can be linear and/or non-linear functions. The max-min formulation is also suitable to handle problems with several explanatory variables ($\mathbf{x} = [\mathbf{x}_1, \mathbf{x}_2, \cdots, \mathbf{x}_p]$). A general recursive procedure can be applied to build segmented models in any circumstance \cite{Bertsekas1975,TZ1981a,TZ1981b}. Let  $u(z) = \max \{0,z\}, ~ \forall z \in \mathbb{R}$. Once $\max \{ a , b \} = a + \max \{ b-a , 0   \} = a + u(b-a)$ for any $a$ and $b$ real, it is possible to join $D$ phases in a convex surface by taking
\begin{eqnarray}
\label{maxmincompleto}
\eta(\mathbf{x},\m{\bm{$\beta$}}) &=& g_1(\mathbf{x},\m{\bm{$\beta$}}_1) + u[g_2(\mathbf{x},\m{\bm{$\beta$}}_2) - ~ g_1(\mathbf{x},\m{\bm{$\beta$}}_1) +  u[ \cdots + \nonumber \\ && u[g_D(\mathbf{x},\m{\bm{$\beta$}}_D) -   g_{D-1}(\mathbf{x},\m{\bm{$\beta$}}_{D-1} )] \cdots ] ] , \nonumber
\end{eqnarray}
Only signs must be changed to join the concave and convex regions together. 

The hierarchical disposal of the max-min operators in Eq. (\ref{maxmincompleto}) is helpful to build piecewise models for multiple regression problems, but it is unnecessary in the context of piecewise-linear models with a single explanatory variable $x$. In this case, the max-min model with $D$ linear regimes is written as follows (appendix \ref{proof1}):
\begin{eqnarray}
\label{abruptomaxmin}
\eta(x | \, \m{\bm{$\beta$}}_1,\m{\bm{$\tau$}},\m{\bm{$\delta$}}) = \alpha_1 + \beta_1 \, x + \sum_{l=1}^{D-1} \delta_l \, u(x-\tau_l), 
\end{eqnarray}
where $\alpha_1$ and $\beta_1$ are the coefficients of the first linear regime, $\delta_l = \beta_{l+1} - \beta_{l}$ and $\tau_l$'s are the change-points between $D\geq2$ linear regimes. This formulation has many advantages since the models are additive, and all the parameters are linear except for the change-points, $\tau_l$'s \cite{Muggeo2008}; furthermore, any prior knowledge about the concavities is required for building models.

Max-min models have discontinuous derivatives at the change-points (or contours), at $\mathbf{x}$ satisfying   
\begin{eqnarray}
g_{l}(\mathbf{x},\m{\bm{$\beta$}}_l) = g_{l+1}(\mathbf{x}, \m{\bm{$\beta$}}_{l+1}), ~~ \forall l \nonumber
\end{eqnarray}
This problem is overcome by replacing the max-min operators for smooth approximating functions. The prevalent idea is to replace the model just in some arbitrary small neighbourhoods of the points where derivatives are discontinuous. The discontinuities are present 
in the model due to the embedded maximum operator $u(z)$, occurring at $z=0$. Thus, smoothing out this operator is sufficient to ensure the existence of derivatives at the change-points, lines, and/or contours. 
Let's consider 
$z_l=z_l(\mathbf{x})=g_{l+1}(\mathbf{x}, \m{\bm{$\beta$}}_{l+1})-g_{l}(\mathbf{x}, \m{\bm{$\beta$}}_l)~~ \forall ~ l.$ The Tishler and Zang's \cite{TZ1981a} method replaces $u(z_l)$ by some smooth function $u_{\gamma}(z_l)$ in the region $-\gamma \leq z_l \leq \gamma$, that is: 
\begin{eqnarray}
\label{paramaxmin}
u_{\gamma}(z_l) = \psi_l(z_l, \gamma) \, I_{\{|z_l| \leq \gamma\}}(z_l) + z_l \, I_{\{z_l>\gamma\}}(z_l), \nonumber
\end{eqnarray}
where $I_{\{\mathbb{A}\}}(z_l)$ is the indicator function defined on the set $\mathbb{A}$ and $\psi_l: \, [-\gamma, \, \gamma] \rightarrow \mathbb{R}$ is a convex, continuously differentiable function of $z_l$ that satisfies the following conditions:  
\begin{itemize}
    \item[(m.1)]{$\psi_l(z_l =  -\gamma, \gamma) = 0   ~~~  \mbox{and}  ~~~  \psi_l(z_l =  +\gamma, \, \gamma) = z_l = \gamma$,}
    \item[(m.2)]{$\partial_{+} \psi_l(-\gamma)  = 0 ~~~~~ \mbox{and} ~~~~~ \partial_{-} \psi_l(+ \gamma) = 1$,}
\end{itemize}
for continuity and differentiability of $u_{\gamma}(z_l)$, respectively. The terms $\partial_{+} \psi_l$ and  $\partial_{-} \psi_l$ are lateral derivatives of 
$\psi_l$ with respect of $z_l$. The parameter $\gamma$ controls the radius of curvature at the phase-transition zone, and the abrupt piecewise-linear model is a special case when $\gamma \rightarrow 0^{+}$. Tishler and Zang suggested the quadratic function $\psi_l(z_l, \gamma_{_l}) = (z_l + \gamma)^2 / (4\gamma)$ as approximation, while Zang provided approximating polynomials of 4$^{th}$, 6$^{th}$ and 8$^{th}$ degrees \cite{Bacon1971,Zang1980}.

Chiu et al. adapted the quadratic Tishler and Zang's approximation for the joining of two straight lines \cite{SW2003,chiu2006}. Their approach is called bent-cable model, and it is given by 
\begin{eqnarray}
\label{chiu}
\eta(x,\m{\bm{$\beta$}}_1,\delta,\tau_1,\zeta) = \alpha_1 + \beta_1 x + \delta \, u_\zeta(x - \tau), ~~~ \zeta>0, \nonumber
\end{eqnarray}
where $\delta = \beta_2 - \beta_1$ and 
\begin{eqnarray}
\label{originalbent}
u_\zeta(x - \tau) = \frac{(x-\tau+\zeta)^2}{4\zeta} \, I_{\{|x-\tau| \leq \zeta\}}(x) + (x-\tau) \, I_{\{x-\tau > \zeta\}}(x). 
\end{eqnarray}
The coefficient $\zeta$ replaces $\gamma$. The parameter $\zeta$ is the half-width of the phase-transition zone and, consequently, controls the radius of curvature of the model at the change-point. 

Khan and Kar noticed many empirical situations in which the original 
quadratic bent is unsuitable \cite{Khan2018}. They generalised the model with a 
more flexible bent, that is given by
\begin{eqnarray}
u_\zeta(x - \tau,k) = \frac{\zeta \, [x -\tau + (k-1) \, \zeta]^k}{(k \, \zeta)^k} 
\, I_{\{ \tau - (k-1)\, \zeta<x\leq \tau + \zeta \}}(x) + (x-\tau) \, 
I_{\{ x-\tau >\zeta \}}(x). \nonumber
\end{eqnarray}
The generalised bent-cable is continuously differentiable for any $k>1$, and the quadratic bent-cable is a special case when $k=2$. This model enables asymmetric 
phase-transition limits in relation to $\tau$. The width of the transition 
region is given by $k \, \zeta$.  

The original bent-cable show discontinuities on second-order derivatives of the log-likelihood function, and the consistency and asymptotic normality for the 
parameter estimators were 
demonstrated under non-standard regularity conditions \cite{chiu2006}. An extension 
of this method is presented in section \ref{SEC-EXT}, which allows the use of 
high-degree polynomials and provides models with first and second-order continuous 
derivatives with respect to $x$ and all the parameters. In this 
way, the standard frequentist procedures for estimation and inference are applicable. 

\section{\label{methodology}Methodology}
The following sections present novel approaches to build smooth piecewise-linear models. 
All the approaches are based on the max-min formulation through Eq. 
(\ref{abruptomaxmin}), by replacement of the abrupt operators. In all 
the cases, the smooth approximations show a close connection 
to specific distributions assumed for a random threshold. 
The methodology is introduced considering two-linear phases, 
but the extension to more than two phases is straightforward.  
\subsection{\label{SEC-EXT}Extended Bent-Cable Models (Ext. BC)}
In the context of linear-piecewise regression with a single explanatory variable, 
all the smooth Tishler and Zang's approximations for $u(z)$ can be rewritten as 
a function of $x$ in the following way   
\begin{eqnarray}
\label{exbent}
u_{\zeta}(x-\tau) = \psi^{*}(x) \, I_{\{|x-\tau| \leq \zeta\}}(x) + (x-\tau) \, I_{\{x-\tau>\zeta\}}(x) \nonumber
\end{eqnarray}
where $\psi^{*}(x) = \psi(z')/|\delta|$, $z' = |\delta| \, (x-\tau)$ and $\zeta =  \gamma / |\delta|$ (section \ref{proof1.1}). The parameter $\zeta$ is the half-width of the phase-transition zone along the $X$-axis, and then the smooth 
approximation takes place in $G = \{x \in \mathbb{R}: \tau-\zeta \leq x \leq \tau+\zeta \}$. The linear phases remain unchanged outside this region.

Conditions of continuity and differentiability (over $x$) must be stated for these models, in order to avoid jumps or kinks at $x=\tau \pm \, \zeta$. The continuity and 
differentiability conditions are given, respectively, by:
\begin{itemize}
\item [(m.1)] $\psi^{*}(x = \tau -\zeta) = 0 ~~~ \mbox{and} ~~~ \psi^{*}(x=\tau +\zeta) = \zeta$, 
\item[(m.2)] $\partial_+ \psi^{*}(x=\tau -\zeta) = 0 ~~~ \mbox{and} ~~~ \partial_{-} \psi^{*}(x=\tau +\zeta) = 1$.  
\end{itemize}
All the Tishler and Zang's approximations, when rewritten as functions of $x$, met 
these conditions (appendix \ref{proof1.6}).

The Chiu's bent-cable (Eq. \ref{originalbent}) is a special case when $\psi(z) = (z+\gamma)^2/4\,\gamma$ (appendix \ref{proof1.2}). Also, a $4^{th}$-degree polynomial bent-cable can be derived from the Tishler and Zang's approximation suggested by Zang \cite{Zang1980} (appendix \ref{proof1.4}):   
\begin{eqnarray}
\label{ex-quartaordem}
\psi^{*}(x) = \frac{\psi(z')}{|\delta|} = -\frac{(x-\tau)^4}{16 \, \zeta^3} + \frac{3\, (x-\tau)^2}{8 \, \zeta} + \frac{(x-\tau)}{2} + \frac{3 \, \zeta}{16}.
\end{eqnarray}
Other alternative is the $6^{th}$-degree polynomial derived from \cite{Bunke1985,Lazaro2001}:
\begin{eqnarray}
\label{sextograu}
\psi^{*}(x) = \frac{(x-\tau)}{2} + \frac{15 \, (x-\tau)^2}{16 \, \zeta} - \frac{5\, (x-\tau)^4}{8 \, \zeta^3} + \frac{3 \, (x-\tau)^6}{16 \, \zeta^5}.
\end{eqnarray} 
Meanwhile the quadratic bent-cable (Eq. \ref{originalbent}) and $4^{th}$-degree polynomial (Eq. \ref{ex-quartaordem}) show a hyperbolic-like shape, the $6^{th}$-degree polynomial bent-cable (Eq. \ref{sextograu}) behaves as a transitional function. 

For the quadratic (original) bent-cable model, the log-likelihood function has discontinuities on the second-order derivatives with respect to $x$, $\tau$ and $\zeta$ along the interception lines, $R_{-} = \{x = \tau - \zeta \}$ and $R_+ = \{x = \tau + \zeta \}$. We see that higher degree polynomials can avoid this problem. Bent-cables 
based on $4^{th}$ and $6^{th}$-degree polynomials have first and second-order continuous derivatives everywhere (with respect to $x$ and all the parameters) when considering a nontrivial $\zeta$ (i.e. $\zeta>0$). It is a desirable characteristic to apply standard asymptotic results  and/or regularisation techniques \cite{Jimenez-Fernandez2016a}.  
\subsection{Random thresholds and smooth transitions}
Structural changes in the model imply the existence of different data generating processes, and smooth transitions between the linear phases are domains in which two or more of these processes take place simultaneously. In this way, the observational unities/sub-unities behave according to distinct phases along the phase-transition zones,  characterising a mixture of processes. The two following approaches assume the existence of random thresholds mediating individual phase-transition events. 
(sections \ref{MM} and \ref{EBCM}).  
\subsubsection{\label{MM}Smooth Transition by State Mixture Model (SMM)}
At first, let's consider that the phenomenon under study is divided into
two linear phases (states), expressed as $g_l(x) = \alpha_l + \beta_l\, x$ and 
$g_{l+1}(x) = \alpha_{l+1} + \beta_{l+1}\, x$, intercepting at $x=\tau$. 
Each experimental/observational unity may be divided into sub-unities 
lying in one of the two possible states ($S = l$ or $S = l+1$), according 
to some probability. 
Let's assume that these probabilities are derived from a continuous 
variable, the random threshold $T$: If $x < T$, the sub-unity behaves according 
to the phase $g_l$ ($S = l$); however, if $x \geq T$, the phase-transition 
$l \rightarrow l+1$ has occurred and the sub-unity behaves according to 
$g_{l+1}$ ($S = l+1$). We assume $f_T(x;\tau,\nu)$ and $F_T(x;\tau,\nu)$ 
as the PDF and CDF for $T$,  
where $\tau$ is the expected value, and $\nu$ is the scale parameter.
In this case, the sub-unities' states $l$ and $l+1$ have the probabilities of 
$P(S=l| \, x) = 1-F_T(x)$ and $P(S=l+1| \, x) = F_T(x)$, respectively. 
Let's consider $Y$ as the sub-unities aggregation response. Thus, the resulting piecewise-linear model is:
\begin{eqnarray}
\eta(x) = \sum_{s=l}^{l+1} E[Y|x,s] \, P(S=s|\, x)
= g_l(x) \, [1-F_T(x)]   + g_{l+1}(x) \, F_T(x), \nonumber
\end{eqnarray}
or in other words, 
\begin{eqnarray}
\label{mixture}
\eta(x)= \alpha_l + \beta_l x + \delta \, (x-\tau) \, F_T(x) 
\end{eqnarray}
where $\delta = \beta_{l+1}-\beta_{l}$. The weights $P(S=l| \, x)$ and 
$P(S=l+1| \, x)$ also can be interpreted as the fractions of sub-unities 
at states $l$ and $l+1$, respectively. 

In this approach, the smooth piecewise model is a 
convex combination of $g_l(x)$ and $g_{l+1}(x)$, and
the function $(x-\tau) \, F_T(x)$ is a smooth 
approximation for the maximum operator $u(x-\tau)$ in Eq. 
(\ref{abruptomaxmin}). Note that the sign formulation 
provides the same result when considering $\trn(x-\tau, \, \gamma) = 2 \, F_T(x) - 1$
(appendix \ref{proof2.1}).

The SMM graph has a transitional shape. Here, we stated the equivalence between the SMMs and transitional models in two steps (Theorem \ref{teo1}). At first, we prove that any non-decreasing transitional function is linked with a random threshold distribution by:
\begin{eqnarray}
\label{fundist}
\trn(x-\tau, \, \gamma)  =  2 F_T(x) - 1, ~~~ -\infty < \tau < +\infty, ~~ \gamma>0, \nonumber
\end{eqnarray}
where $F_T(x)$ is its CDF (appendix \ref{proof2.2}). Then, we proved that the SMM approach produces transitional models when the random threshold $T$ has: (a) continuous PDF and (b) integrable PDF ($E|T|<+\infty$) or limited support (appendix \ref{proof2.3}). 
\begin{theorem}
\label{teo1}
A piecewise-linear model based on a smooth, non-decreasing transitional approximation is equivalent to the SMM for that $T$ is distributed according to  
\begin{eqnarray}
\label{underdist}
f_{T} (x | \tau, \, \gamma) = \frac{1}{2} \frac{d}{d x} \trn(x-\tau , \gamma). 
\end{eqnarray}
\end{theorem}
The hyperbolic tangent $\tanh(.)$, for example,  
is naturally linked with the Logistic distribution 
for $T$ (appendix \ref{proof2.4}). Furthermore, some SMMs based on distributions with 
limited support are related to the Ext. BC through $\psi^{*}(x) = (x-\tau) \, F_T(x)$. In other words, there are situations for which $F_T(x)$ satisfies
\begin{eqnarray}
F_T(x) = \frac{\psi^*(x)}{x-\tau} ~~~~~ \mbox{and} ~~~~~ \trn(x-\tau) = 2 \, \frac{\psi^*(x)}{x-\tau} -1, ~~~~ \tau -\zeta \leq x \leq \tau+\zeta. \nonumber
\end{eqnarray}
In these cases, the SSM and Ext. BC approaches are equivalent. 
For example, the high-degree polynomial approximation \cite{Bunke1985,Lazaro2001} 
in Eq. (\ref{sextograu}) is linked with the bi-weight (or quartic) kernel function (section \ref{proof2.5}). This model will be called the Quartic bent-cable (Q-BC) hereinafter. 

New transitional models can be obtained from Eq. (\ref{underdist}) when considering any integrable, continuous distribution for $T$. In this context, nothing can be said about non-integrable distributions. 
However, we see that Cauchy's distribution characterises an example 
for which a transitional model cannot be derived (appendix \ref{proof2.6}).

\subsubsection{\label{EBCM}The Expected Bending-Cable Models (Exp. BC) }
In the SMM approach, the smoothing of the operator $u(x-\tau)$ considers 
$\tau$ as a fixed value. It disregards 
the change-point randomness when smoothing abrupt operators, 
and the linear phases are artificially connected causing the 
`bulging phenomenon', which is uncommon in nature.

For correction, we suggest the smoothing of $u(x-T)$ instead. It is 
equivalent to the model marginalisation over $T$ distribution when 
considering the intercept and slopes as fixed (appendix \ref{proof3.1}):    
\begin{eqnarray}
\label{expectedbent}
E[Y|x] = \alpha_l + \beta_l \, x + \delta_l \,E[u(x-T)],    
\end{eqnarray}
where
\begin{eqnarray}
\label{expectedmax}
E[u(x-T)] = (x - \tau) F_T(x) + \int_{x}^{+\infty} (T - \tau) \, dF_T(T). 
\end{eqnarray}
Let's consider each observational unity $Y$ formed by $M$ sub-unities $Y_m$, for which the iid random thresholds are given by $T_m$'s. 
Then,  
\begin{eqnarray}
E \Big [\sum_m Y_m|x \Big ] &=& M\,a + M\,b\, x + d \, E \Big [\sum_m u(x-T_m) \Big ] = M \Big \{ a + b\, x + d \, E[u(x-T)] \Big \}, \nonumber  
\end{eqnarray}
where $M \, a = \alpha_l$, $M \, b = \beta_l$, and $M \, d = \delta_l$. Thus, this conditional response is also true for aggregated responses.   

Note that the model in Eq. (\ref{expectedbent}-\ref{expectedmax}) is equivalent to the model obtained by replacing the modulus in Eq. (\ref{sinal}) with 
{\small{
\begin{eqnarray}
\label{expectedsignal}
E|x-T|=(x-\tau) \, \Big [ 2 \, F_T(x) - 1 \Big ] + \int_{x}^{\infty} (T-\tau) \, dF_T(T) - \int_{-\infty}^{x} (T-\tau) \, dF_T(T). 
\end{eqnarray}
}}
Hereinafter, these models will be called Expected bending-cables (Exp. BC). 

For appropriate asymptotic behaviour of the model, the terms 
\begin{eqnarray}
\int_{x}^{+\infty} (T - \tau) \, dF_T(T) ~~~ \mbox{and} ~~~ \int_{-\infty}^{x} (T-\tau) \, dF_T(T)   \nonumber
\end{eqnarray}
must go to zero at infinity, and that requires an integrable $T$ random threshold ($E|T|<+\infty$; appendix \ref{proof3.2-2}).  
  
As previously seen, the continuity of first and second-order derivatives with 
respect to $x$ is a desirable characteristic. By applying the Leibniz's 
integral rule, it is easy to see that
\begin{eqnarray}
\frac{d}{dx} E[u(x-T)]  =  F_T(x) ~~~~ \mbox{and}  ~~~~ \frac{d^2}{dx^2} E[u(x-T)]  =  f_T(x). \nonumber
\end{eqnarray}
So, for any continuous PDF, the Exp. BC has continuous 
first and second-order derivatives over $x$. In this model, the sigmoid 
function $F_T(x)$ replaces the jump on the first-order derivative of $u(.)$. 

The Exp. BC graph has a hyperbolic-like shape. Any Exp. BC model  
can be rewritten as a valid hyperbolic approximation (appendix \ref{proof3.3}), as we stated in the following theorem:
\begin{theorem}
\label{teo2}
The Exp. BC models are smooth piecewise-linear hyperbolic models, for which the hyperbolic approximation satisfying (h.1 - h.4) is given by:
\begin{eqnarray}
\label{hypteo}
\hyp(x-\tau, \gamma) =  2 \, F_T(x) - 1 - 2 \frac{\int_{-\infty}^{x} (T-\tau) \, dF_T(T)}{x-\tau} \nonumber.
\end{eqnarray}
Any integrable $T$ with a continuous PDF has an associated hyperbolic approximation given in this form. 
\end{theorem}

In this case, the $\hyp(.)$ is a transitional function minus a correction factor (CF), that is $\hyp(x-\tau, \gamma) = \trn(x-\tau, \gamma) - CF$. For the Exp. BC models we see that $ \frac{d}{dx} [ (x-\tau) \, \hyp(x-\tau, \gamma) ]  = 2\, F_T(x) - 1$, then some hyperbolic models found in the literature are related to specific distributions for $T$ by the following equation: 
\begin{equation}
F_T(x) =  \frac{d}{dx} \Big [ \frac{(x-\tau) \, \hyp(x-\tau, \gamma)}{2} \Big ] + \frac{1}{2}. \nonumber     
\end{equation}
For example, the hyperbolic model \cite{Griffiths1973} is related to 
the non-standard t-distribution with 2 degrees of freedom (appendix \ref{prooffinal1}). 
Moreover, the $\lne(.)$  
\cite{Jimenez-Fernandez2016a} is equivalent to the Logistic distribution 
for $T$ (appendix \ref{prooffinal2}). 

The Exp. BC models based on distributions with limited support are also 
related to the Ext. BC models in the following way
\begin{eqnarray}
\label{verify}
\frac{d}{dx} \psi^{*}(x) = \frac{d}{dx} E[u(x-T)] = F_T(x) , ~~~  \tau - \zeta < x < \tau + \zeta. 
\end{eqnarray}
That is, the approaches are equivalent if the Eq. (\ref{verify}) holds for some CDF.  
For example, the Chiu's bent-cable \cite{chiu2006} is related to a Uniform $T$  
(appendix \ref{outra}), and the discontinuities at $x=\tau \pm \zeta$ for the second-order derivatives are justified by the discontinuities of the Uniform PDF. Furthermore, the $4^{th}$-degree polynomial bent-cable in the Eq. (\ref{ex-quartaordem}) is linked with the Epanechnikov's distribution for $T$ (appendix \ref{proof3.5}). This model will be called the Epanechnikov's bent-cable (E-BC) hereinafter.   

The Khan and Kar's  \cite{Khan2018} generalised bent-cable is not an Ext. BC. But, by making $\frac{d}{dx}E[u(x-T)] = F_T(x)$ for $\tau - (k-1)\, \zeta<x\leq \tau + \zeta$, we proved that the Khan and Kar's model is an Exp. BC when $T$ has the Exponentiated Uniform distribution \cite{Ramires2019} (appendix \ref{proofunif}).

\subsection{\label{SEC-INTER}New models and inter-relationships}
In the study, we introduced two high-degree polynomial bent-cables 
as adapted versions of Zang's and Bunke and Schulze's smooth 
approximations: the Epanechnikov's bent-cable (E-BC) in Eq. (\ref{ex-quartaordem}), and 
the Quartic bent-cable (Q-BC) in Eq. (\ref{sextograu}). These models provide a clear 
delimitation of the phase-transition zone, that is $G = [\tau-\zeta, \tau+\zeta]$. 
Furthermore, they are linked to specific distributions for the random threshold $T$ 
- Epanechnikov and Bi-weight (or quartic), respectively. These models are suitable alternatives to the Chiu's bent-cable since 
they provide continuous first and second-order derivatives everywhere. 
The Q-CB model has the transitional shape, seen by many authors as undesirable. 
However, as any transitional model, it can be corrected following the  
Eq. (\ref{hypteo}), resulting in a Q-BC hyperbolic model (not shown).     

Our methodology allows us to build the transitional 
and hyperbolic versions of the piecewise-linear model for any well-behaved random distribution for $T$. The Normal model seems a natural choice; alternatively, in a more generic framework, we would consider the Skew-Normal distribution instead, in order to accommodate asymmetry in the phase-transition process. The Skew-Normal PDF is given by
\begin{eqnarray}
f_T(x| \xi, \gamma, \lambda) = \frac{2}{\gamma \, \sqrt{2\, \pi}} \, e^{-\frac{(x-\xi)^2}{2 \, \gamma^2} } \, \int_{-\infty}^{\lambda \,  ( \frac{x-\xi}{\gamma} )} f_Z(z) \, dz, \nonumber
\end{eqnarray}
where $\xi$, $\gamma$ (positive) and $\lambda$ are the location, scale, and shape parameters, respectively. Also, the variable $Z$ has standard normal distribution 
and $f_Z(z)$ is its density. The change-point $E(T) = \tau$ is given by 
\begin{eqnarray}
\tau = \xi + \gamma \, \lambda \, \sqrt{ \frac{2}{\pi \, (1+\lambda)^2}}. \nonumber
\end{eqnarray}
The Exp. BC models based on the Skew-Normal and Normal distributions 
will be called Normal bent-cable (N-BC) and Skewed-Normal bent-cable (SN-BC), respectively. The SN-BC has three nonlinear (bent) parameters, increasing the complexity of estimation and inference, but the N-BC is nested within it, and thus the asymmetry on the bent can be tested easier. 

Models based on distributions with unlimited support, such as the N-BC and SN-BC, have no clear distinction between the linear phases and phase-transition zones. However, the fitted curves provide values for the PDF parameters and, thus, the underlying distribution is drawn. So, we recommend the prior specification of quantiles (2.5\% and 97.5\% percentiles, for example) to define the phase-transition zones. 

All models that we found in the literature fits one of our approaches (Table \ref{tab01});  and, therefore, they have probabilistic interpretations in 
the light of random thresholds. 

\begin{table}[p]
\tbl{Interrelationships between the smooth piecewise-linear models in the light of random thresholds. Modelling approaches: Extended Bent-Cable (Exp. BC), State Mixture Modelling (SMM), and Expected Bending-Cable (Exp. BC). Models: hyperbolic tangent (Tanh) \cite{Bacon1971}. Log and Exponential (Lne) \cite{Jimenez-Fernandez2016a}. Log hyperbolic cosine (Lch) \cite{Lazaro2001}. Hyperbolic model (Hyp)  \cite{Griffiths1973,Watts1974}. Bent-Cable (BC) \cite{chiu2006}. Generalised Bent-Cable (G-BC) \cite{Khan2018}. Epanechnikov Bent-Cable (E-BC) adapted from Zang \cite{Z1980}. Quartic Bent-Cable (Q-BC) adapted from \cite{Bunke1985,Lazaro2001}. Normal Bent-Cable (N-BC) and Skew-Normal Bent-Cable (SN-BC).}
{\begin{tabular}{llcccc} \toprule
\em Model &\em Shape &\em \hspace{0.5cm} Ext. BC \hspace{0.5cm} &\em \hspace{0.5cm} SMM \hspace{0.5cm} &\em \hspace{0.5cm} Exp. BC \hspace{0.5cm} &\em \hspace{0.5cm} Threshold distribution \hspace{0.5cm}\\
\hline
Tanh & Transitional & No & Yes & No  & Logistic \\
Lne  & Hyperbolic & No & No & Yes & Logistic \\
Lch  & Hyperbolic \textsuperscript{a} & No & No & Yes & Logistic \\
Hyp  & Hyperbolic  & No & No  & Yes & Non-standard t\\
BC   & Hyperbolic &  Yes & No & Yes & Uniform \\
G-BC & Hyperbolic &  No  & No & Yes & Exponentiated Uniform \\
E-BC & Hyperbolic &  Yes & No & Yes &  Epanechnikov \\
Q-BC & Transitional &  Yes & Yes & No & Bi-weight\\
N-BC & Hyperbolic  &  No & No & Yes & Normal\\
SN-BC & Hyperbolic &  No & No & Yes & Skew-Normal\\ \bottomrule
\end{tabular}}
\tabnote{\textsuperscript{a}Lch is just a vertical displacement of the Lne approximation.}
\label{tab01}
\end{table}

\subsection{\label{estimation}Estimation and Inference}
We considered only hyperbolic models (BC, E-BC, N-BC, and SN-BC) since the data don't show evidence for any `bulge' at the transition.  Despite that, the methods employed here are fully applicable to any smooth piecewise-linear model. Models were fitted by maximum likelihood (ML) estimation, considering that:   
\begin{eqnarray}
Y_i  =  \eta(x_i | \, \bm{\theta}) + E_i,  ~~~ i = 1 \cdots n. \nonumber
\end{eqnarray}
We assumed normal, independent, and equally 
dispersed observations ($E_i \sim N(0 , \, \sigma^2 \, \bm{I})$). 
The parameter vector 
for smooth piecewise-linear models is 
given by $\bm{\theta}^T = (\bm{\beta}^T, \bm{\phi}^T)$, 
where $\bm{\beta}^{T} = ( \alpha_l, \, \beta_l, \, \delta_l)$ is the 
vector of linear parameters, while the vector of non-linear bent-parameters 
$\bm{\phi}$ is given by: $\bm{\phi}^{T} = (\tau, \zeta)$ for BC and E-BC models; 
$\bm{\phi}^{T} = (\tau, \gamma)$ for N-BC; 
and $\bm{\phi}^{T} = (\xi, \gamma, \lambda)$ for SN-BC model. 

The $\hat{\bm{\theta}}$ was that maximised 
\begin{eqnarray}
\label{log-lik}
\log L (\bm{\theta}) = - \frac{n}{2} \log(2\, \pi) 
- \frac{n}{2} \log(\sigma^2) - \frac{1}{2 \, \sigma^2} \sum_{i=1}^{n}
(y_i - \eta(x_i| \, \bm{\theta}))^2. \nonumber
\end{eqnarray}
The $\hat{\sigma^2}$ was obtained afterwards, by making
\begin{eqnarray}
\hat{\sigma^2} =  \frac{1}{n} \sum_{i=1}^{n} (y_i - \eta(x_i| \, \hat{\bm{\theta}})^2. \nonumber
\end{eqnarray}
The model non-linearity requires iterative procedures starting from initial guesses. The smoothing parameter ($\gamma$ or $\zeta$) often starts from an arbitrarily small value. In this study, its initial value was fixed as $\gamma^{(0)} = .005 \, \Delta$, where $\Delta = x_{(n)} - x_{(1)}$. We conducted a grid search based on 100 candidates for $\tau$, equally spaced over the $x$-range, in order to get starting values for $\tau$ and $\bm{\beta}$. They were chosen $\tau^{(0)}$ and $\bm{\beta}^{(0)}$ which maximised the log-likelihood function among these 100 possible fittings of the abrupt model. Note that the estimates of $\bm{\beta}^{(0)}$ are promptly determined by Ordinary Least Square (OLS) equations when $\tau^{(0)}$ is known. 

We used the Genetic Algorithm (GA) to fit the models. It has the advantage of starting from a population of potential solutions (chromosomes), exploring more effectively the search space and avoiding to get stuck in local optima. In the GA, chromosomes are ranked according to their fitness (log-likelihood values), and the better ones reproduce more often. At every step, the algorithm stores copies of the best chromosomes to create the next generation. After that, the chromosomes are combined, peer to peer, and pieces of information (genes) are randomly selected and changed by other plausible values, in a process that mimics the crossover and genetic mutation. It results in the offspring - that is, the next generation of potential solutions. Thus the population gradually improves, step by step, until the fitness function stagnation, when the search procedure stops and the best chromosome is selected as the solution. 

In our study, the population size was fixed in 100 chromosomes, and the algorithm employed at least $5,000$ generations to get the solution. 
The starting population included five replicates of the initial guesses; and the remaining 95 chromosomes encompassed randomly selected values from the search space, i.e. the Cartesian product of: $\beta^{(0)} \pm 5.2 \, |\beta^{(0)}|$ for any parameter of $\bm{\beta}$; $\tau^{(0)} \pm 0.2 \, \Delta$ for $\tau$; from $0.005 \, \Delta$ up to  $0.2 \, \Delta$ for $\gamma$; from $-30.0$ up to $30.0$ for $\lambda$.  

Irregular likelihood surface is a common problem in piecewise-linear regression \cite{SW2003, chiu2006}. Then, the inspection of the likelihood (or deviance) 
surface is a crucial step in the validation of classical inferences. 
We explored the deviance surfaces over $(\tau,\gamma)$ 
or $(\tau,\gamma)$-spaces, in the neighbourhoods of the ML estimates 
$\hat{\bm{\theta}} = [\hat{\bm{\beta}},  \, \hat{\bm{\phi}}]^{T}$.
The domain was replaced by a grid of 40$\times$40 points, and OLS 
estimates for $\bm{\beta}$ were calculated for each pair $(\tau,\gamma)$
or $(\tau,\zeta)$ in the grid. 

Note that the ML and LS estimates for $\bm{\beta}$ are equivalent under the assumption of Normal errors, and both minimise the sum of squares as follows:     
\begin{equation} 
SS(\bm{\beta}, \bm{\phi}) = \sum_{i=1}^{n} (y_i - \eta(x_i))^2. \nonumber 
\end{equation}
For a given set of bent-parameters $\bm{\phi}$, the problem of obtaining OLS estimates 
for $\bm{\beta}$ is solved promptly by simple matrix algebra, by considering 
\begin{equation}
SS(\bm{\beta} | \bm{\phi}) = (\bm{y} - \mathbf{X} \bm{\beta} )^T \, (\bm{y} - \mathbf{X} \bm{\beta} ), \nonumber   
\end{equation}
where the model matrix is expressed by 
\begin{equation}
\mathbf{X} = \left [
\begin{array}{ccc}
1  & x_1 & T(x_1, \bm{\phi}) \\
1  & x_2 & T(x_2, \bm{\phi})  \\
\vdots & \vdots &  \vdots \\ 
1  & x_n & T(x_n, \bm{\phi}) 
\end{array}
\right ]. 
\end{equation}
The functions $T$ are the bending/transitional approximations, which depends on 
the chosen method: 
\begin{equation}
T(x, \bm{\phi}) = \left \{
\begin{array}{ll}
u_{\zeta}(x-\tau),   & \mbox{for the Ext. BC}\\
(x-\tau) \, F_{T}(x),  &  \mbox{for the SMM}\\
(x-\tau) \, F_{T} (x) + \int_{x}^{+\infty} (T-\tau) \, dF_T,  &  \mbox{for the Exp. BC}. 
\end{array} \right .   \nonumber
\end{equation}
Thus, the OLS estimates for $\bm{\beta}$ are given by
\begin{equation}
\bm{\beta}^* = (\mathbf{X}^T \mathbf{X})^{-1} \,\mathbf{X}^T \mathbf{y}.
\end{equation}
The deviance drop over $(\tau, \gamma)$-space was calculated as follows
\begin{eqnarray}
D(\tau, \gamma) = - 2 \Big [\log L(\hat{\bm{\theta}}) - \log L( \bm{\beta}^* | \tau, \gamma) \Big ] ,  \nonumber 
\end{eqnarray}
where $\bm{\beta}^*$ is the OLS estimates when considering the pair $(\tau, \gamma)$ or $(\tau, \zeta)$ as fixed at prior.

In this case, the Wilk's statistic for the likelihood ratio test is given by $LRT = - D$. LRT is asymptotically distributed according to $\chi^2_{(2)}$, and therefore, the ellipsoidal 95\% confidence region for $(\tau,\gamma)$ can be approximated by 
$\{(\tau,\gamma) \in (-\infty,+\infty)\times[0,+\infty) : D(\tau, \gamma) > - 5.99 \}$. However, this approximation often presents poor coverage, and its reliability depends 
on a regular, parabolic surface for $D$ \cite{chiu2006}. We employed the LRT for 
testing the SN-BC against N-BC. The hypotheses were $H_0: \lambda = 0$ vs. $H_A: \lambda \neq 0$; thus, the rejection of the null hypothesis would imply an asymmetrical bent-cable. 

The fit diagnostic was carried out by residual plots and normality tests. 
Furthermore, the lack of fit was tested by the F-statistic. We compared the 
models, in terms of plausibility, by using the relative likelihood criterion. 
The use of this criterion is justified by a quite regular 
likelihood surface that was drawn from Cook's data. The 
plausibility of a model in a set of $M$ predefined alternatives 
is given by
\begin{eqnarray}
Pr(\mbox{model}~m) = e^{(AIC_{min}-AIC_m)/2}, ~~~ m=1, \cdots, M,\nonumber 
\end{eqnarray}
which is based on the Akaike's Information Criterion (AIC). 

\subsection{\label{application}Application}
We applied the bent-cable models to describe the R.A. Cook's data on stagnant 
surface layer height as a function of the rate flow of water (and suspended particles) down an inclined channel, under the action of a surfactant. These data are present in the Cook's doctorate thesis at Queen's University, and they were published by Bacon and Watts to illustrate the use of their transitional model, based on the hyperbolic tangent as an approximation for $\sgn(x-\tau)$ \cite{Bacon1971}. Cook's data don't show any evidence for `bulge' formation, and for this reason, many authors revised the analyses using hyperbolic models, such as the Griffiths and Miller's model and the Chiu's bent-cable \cite{SW2003,chiu2006}. The previous studies showed well-behaved fits for this data set, with approximated parabolic log-likelihood profiles at the neighbourhoods of the ML estimates \cite{chiu2006}.     

Chiu et al. attempted to answer if a smooth model could be considered more 
plausible than the abrupt model (or broken-line regression). For the Cook's data, 
they found  an overwhelming evidence against the abrupt model, concluding that 
the degree of smoothness should be incorporated in analyses - being estimated or given at prior \cite{chiu2006}. Our study is a complement, and it aimed to answer 
the following two questions: (a) An asymmetrical bent-cable could provide a better description for the Cook's data? (b) It is possible to choose among bent-cable alternatives or, similarly, to specify the underlying random threshold 
distribution?        

When comparing the SN-BC and N-BC, the LRT was $\Lambda = 0.0230$ ($P = 0.879$, $\chi^2_{(1)}$ test). Thus the N-BC model was more parsimonious, indicating no 
need to incorporate asymmetry on the bent-function. 
The shape parameter for SN-BC was $\hat{\lambda} = -0.818$, 
not differing statistically from zero.

Concerning the second question, all the symmetric bent-cable models (BC, E-BC and N-BC) performed well, being plausible according to their relative likelihood (Tab. \ref{tab02}). They have the same number of parameters (five), but Chiu's bent-cable provided the smallest AIC. The parameter estimates were $\hat{\alpha} =0.569$, $\hat{\beta} = - 0.398$, $\hat{\delta} = - 1.064$, $\hat{\tau} = - 0.056$,
and $\hat{\zeta}=0.428$. 

\begin{table}[p]
\tbl{Comparison of smooth piecewise-linear models in the description of R.A. Cook's data on stagnant layer surface height \cite{Bacon1971,SW2003}. Models: Bent-Cable (BC)  \cite{chiu2006}. Epanechnikov Bent-Cable (E-BC) adapted from Zang \cite{Z1980}. Normal Bent-Cable (N-BC), and Skew-Normal Bent-Cable (SN-BC). Models compared from 
log-likelihood, Akaike's Information Criterion (AIC), and relative likelihood. }
{\begin{tabular}{lccc} \toprule
\em Model \hspace{2cm} & \em \hspace{1cm} Log-Likelihood \hspace{1cm} &\em  \hspace{1cm} AIC  \hspace{1cm} &\em \hspace{1cm} $Pr(\mbox{model}~j)$ \hspace{1cm} \\
\hline
BC    & 85.03048 & -158.061  & 1.000\\
E-BC  & 84.95691 & -157.914  & 0.929\\
N-BC  & 84.66888 & -157.338  & 0.697\\
SN-BC & 84.68381 & -155.368 & 0.260\\ \bottomrule
\end{tabular}}
\label{tab02}
\end{table}

\begin{figure}[p]
    \centering
    \includegraphics[width=14cm,height=12cm]{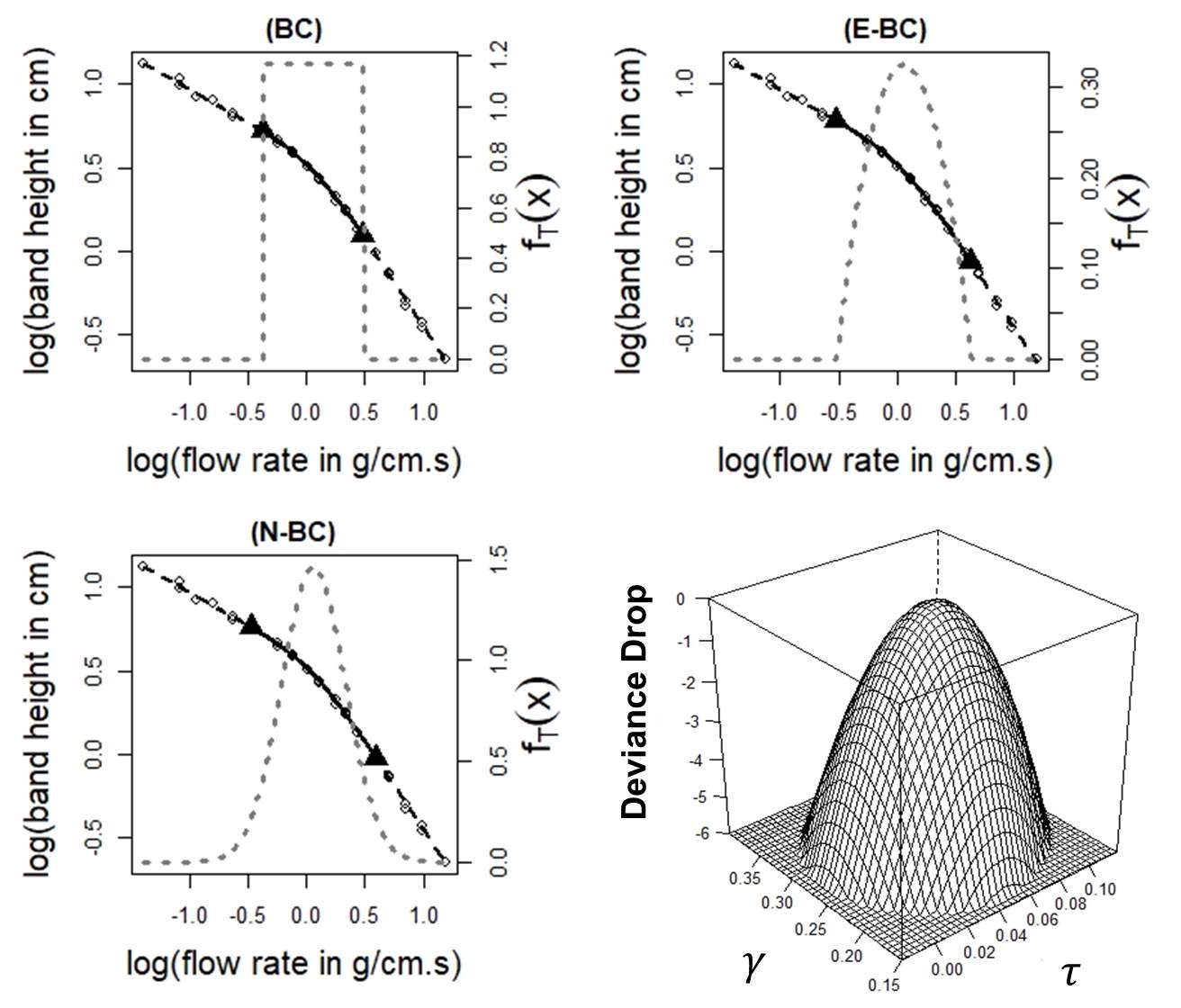}
    \caption{(a) Fittings of smooth piecewise-linear models to the R.A. Cook's data on stagnant layer surface height \cite{Bacon1971,SW2003}. Models: Bent-Cable (BC) \cite{chiu2006}. Epanechnikov Bent-Cable (E-BC) adapted from Zang \cite{Z1980}. Normal Bent-Cable (N-BC). Experimental data (circles), linear phases (dotted lines) and 
    transition zones (continuous lines). The black triangles limits separate the 
    linear phases and transition zones. (b) The log-likelihood deviance surface 
    for $\tau$ and $\gamma$ with respect to the best fit of SN-BC model. All deviance values above -5.99 are consistent with the data at 95\% confidence level - based on the $\chi^2_{(2)}$ approximation.}
    \label{fig:my_label}
\end{figure}

The experimental design shows replicated $x$-values, allowing to perform the F-test for lack of fit. All the models were fitted well, without evidence for the lack of fit ($P>0.05$). The fitted curves and parameter estimates allow us to draw the underlying distribution for $T$ and to determine the phase-transition zones (Fig. \ref{fig:my_label}). Chui's model provided the narrower transition zone, $x \in [-0.373, 0.484]$. Results were identical to that described by Chiu et al. \cite{chiu2006}. 

In summary, for the two questions raised in our study, the answer was `no': there was no evidence for asymmetrical bent-function; and we could not differentiate among bent-cable alternatives, to choose the better description and/or appropriate threshold distribution.       

We observed that points within the transition zones can be highly influential. For small sample sizes, little changes in the y-values lead to great changes in the bent-cable graph. Maybe the test of asymmetry and comparisons between bent-cable formulations require more expressive sample sizes.  

\section{Conclusion}
This study introduced a unifying framework for the smooth piecewise-linear modelling. 
Our methodology encompassed seven models commonly found in the literature; furthermore, we introduced three new bent-cable models: the Epanechnikov, Normal and Skewed-Normal Bent-Cables. We also derived a formal test to asymmetry on the bent-function, by comparing the Normal and Skewed-Normal Bent-Cables. Our methodology was applied to describe the R.A. Cook's data on stagnant surface layer height. We conclude that the transition between linear phases is gradual and symmetrical for these data; furthermore,
all the symmetrical bent-cable models that we tested can be considered a good description of the process.

\section*{Funding}
We want to thank the Brazilian National Council for Scientific and Technological Development (CNPq) for the research grant (Program GM/GD - Nº 140518/2011-8). 	

\bibliographystyle{tfnlm} 
\bibliography{article}

\appendix
\section{\label{proof1}Max-min general piecewise-linear formulae}
Let $D\geq2$ linear regimes expressed by $g_j(x | \, \m{\bm{$\beta$}}_j) = \alpha_j + \beta_j \, x$ for $j=1, \cdots, D$. The continuity constrains determine that $\tau_l  = (\alpha_l - \alpha_{l+1})/(\beta_{l+1} - \beta_{l}), ~l=1,\cdots, D-1$. 

Initially, consider the problem of joint two linear phases ($l$ and $l+1$) at the change-point $\tau_l$. If $\beta_{l+1} > \beta_{l}$, then the response curve has a convex graph and the regimes are joined by using the $\max$ function in the
following way
\begin{eqnarray}
\eta(x | \, \m{\bm{$\beta$}}) &=& \max \{g_1(x | \,\m{\bm{$\beta$}}_1),g_2(x,\m{\bm{$\beta$}}_2)\} \nonumber \\  
&=& g_1(x | \,\m{\bm{$\beta$}}_1) + \max \{g_2(x | \, \m{\bm{$\beta$}}_2)- g_1(x | \,\m{\bm{$\beta$}}_1), 0\} \nonumber \\
&=& \alpha_l + \beta_l x + u(z_l). \nonumber
\end{eqnarray}
On the other hand, if $\beta_{l+1} > \beta_{l}$, the graph is concave and the regimes are joined by means of the $\min$ function, as shown
bellow
\begin{eqnarray}
\eta(x | \,\m{\bm{$\beta$}}) &=& \min \{g_1(x | \,\m{\bm{$\beta$}}_1),g_2(x | \, \m{\bm{$\beta$}}_2)\} 
\nonumber \\ 
&=& - \max \{ - g_1(x | \, \m{\bm{$\beta$}}_1), - g_2(x 
| \, \m{\bm{$\beta$}}_2)\} \nonumber \\
&=& \alpha_l + \beta_l x -  u(-z_l). \nonumber
\end{eqnarray}
The case in which $\beta_{l+1} = \beta_{l}$ is trivial, given that there is no phase change in this situation.

More regimes can be added to current model, recursively. Thereby, after the addition of all regimes, the model is given by
\begin{eqnarray}
\label{modelomaisoumenos}
\eta(x | \, \m{\bm{$\beta$}}) = \alpha_1 + \beta_1 x  + \sum_{l=1}^{D-1} \pm ~ u(\pm z_l). \nonumber
\end{eqnarray}
The signs to be used depend on the concavities at the change-points. A simple way to express the piecewise-linear models without worrying about the concavities is given by the following reparameterization: consider $\delta_l = \beta_{l+1} - \beta_{l}$, in such a way that $z_l = \alpha_{l+1} - \alpha_{l} + \delta_l x$. Then, it is verified that
$\alpha_{l+1} - \alpha_{l} = -\delta_{l} \tau_l$ and $z_l = \delta_l  (x - \tau_l)$. Consequently, if the graph is convex in $x=\tau_l$, then
$\beta_{l+1} > \beta_{l} \Leftrightarrow \delta_l > 0$ and, therefore, $u(z_l) = u(\delta_l  (x - \tau_l)) = \delta_l  u(x - \tau_l)$; on the other
hand, if the graph is concave in $x=\tau_l$, then $\beta_{l+1} > \beta_{l} \Leftrightarrow \delta_l > 0$ and 
$- u( - z_l) = - u( - \delta_l  (x - \tau_l)) = \delta_l  u(x - \tau_l)$. Thus, the model is expressed simply by
\begin{eqnarray}
\label{nome1}
\eta(x | \, \m{\bm{$\beta$}}_1,\m{\bm{$\tau$}},\m{\bm{$\delta$}}) = \alpha_1 + \beta_1 \, x + \sum_{l=1}^{D-1} \delta_l \, u(x-\tau_l), \nonumber
\end{eqnarray}
where $\m{\bm{$\delta$}}^{T} = (\delta_1, \cdots, \delta_{D-1})$. This formulation is commonly used these days in piecewise regression \cite{Muggeo2008}, but we are not sure if any author recognised it before as a special case of the max-min models. 

\section{The extended bent-cable model}
\subsection{\label{proof1.1} Model derivation}
{\it For any Tishler and Zang's approximation $u_{\gamma}(z')$, there is an approximation $u_{\zeta}(x-\tau)$ which gives the same smooth piecewise-linear model.} \\

\noindent Let's consider two linear phases $g_l(x) = \alpha_l + \beta_l\, x$ and 
$g_{l+1}(x) = \alpha_{l+1} + \beta_{l+1}\, x$ to be joined. Note that $z = (\beta_{l+1} - \beta_{l}) \, x + \alpha_{l+1} - \alpha_l = \delta \, (x - \tau)$, where 
$\delta = \beta_{l+1} - \beta_{l}$ and $\tau=(\alpha_l - \alpha_{l+1})/(\beta_{l+1} - \beta_{l})$. Also, note that:  
\begin{eqnarray}
\delta \, u(x-\tau) = \left \{ \begin{array}{ll}
 u(z) & \mbox{if} ~~~ \beta_{l+1} > \beta_{l} \Leftrightarrow \delta > 0, \\
 - u(-z) & \mbox{if} ~~~ \beta_{l+1} < \beta_{l} \Leftrightarrow \delta <0.
\end{array}
\right . \nonumber
\end{eqnarray}
In short, it can be expressed as $u(x-\tau) = u(z')/|\delta|$, 
where $z' = z \, \sgn(z) = |\delta| \, (x-\tau)$. Thus, 
by analogy, we can make $u_{\zeta}(x-\tau) = u_{\gamma}(z')/|\delta|$
for any non-trivial case ($\delta \neq 0$), obtaining  
\begin{eqnarray}
\psi^{*}(x) \, I_{\{|x-\tau| \leq \zeta\}}(x) + (x-\tau) \, I_{\{x-\tau>\zeta\}}(x) = \frac{
\left [ \psi(z') \, I_{\{|z'| \geq \gamma\}}(z') + z' \, I_{\{z'>\gamma\}}(z') \right ]} { |\delta|}. \nonumber
\end{eqnarray}
Since $z'/|\delta| = x-\tau$, then we have $\psi^{*}(x) = \psi(z')/|\delta|$ which takes place 
in $\{ -\gamma \leq z' \leq +\gamma \}$ and, therefore, $\zeta =  \gamma / |\delta|$. 

\subsection{\label{proof1.6}Continuity and differentiability conditions}
{\it For any Tishler and Zang's approximation, the derived (extended) bent-cable $u_{\zeta}(x-\tau)$ met the continuity and differentiability conditions} \\

\noindent Let $\delta \neq 0$, thus
\begin{eqnarray}
\psi^{*}(x = \tau+\zeta) = \psi(|\delta| \, \zeta)/|\delta| = \psi(+\gamma)/|\delta| = +\gamma/ |\delta| = + \zeta \nonumber
\end{eqnarray}
and 
\begin{eqnarray}
\psi^{*}(x = \tau-\zeta) = \psi( -|\delta| \, \zeta)/|\delta| = \psi(-\gamma)/|\delta| = 0 \nonumber.
\end{eqnarray}
Then the continuity restrictions are met.  Since the $\psi^{*}(x)=\psi(z')/|\delta|$, the differentiability is proved by 
\begin{eqnarray}
\frac{\partial \psi^{*}(x)}{\partial x}  = \frac{1}{|\delta|} \frac{\partial \psi(z')}{\partial z'}    
\frac{\partial z'}{\partial (x-\tau)} \frac{\partial (x-\tau) }{\partial x} = \frac{\partial \psi(z')}{\partial z'}, \nonumber
\end{eqnarray} 
since it implies lateral derivatives with the same value, that is:
\begin{eqnarray}
\lim_{x \rightarrow (\tau-\zeta)^{+}} \frac{\partial \psi^{*}(x)}{\partial x} = \lim_{z'\rightarrow -\gamma^{+}}\frac{\partial \psi(z')}{\partial z'} = 0, \nonumber
\end{eqnarray}
and
\begin{eqnarray}
\lim_{x \rightarrow (\tau+\zeta)^{-}} \frac{\partial \psi^{*}(x)}{\partial x} = \lim_{z'\rightarrow +\gamma^{-}}\frac{\partial \psi(z')}{\partial z'} = 1. \nonumber
\end{eqnarray}

\subsection{\label{proof1.2} The Chiu's bent-cable}
{\it
The original (quadratic) bent-cable \cite{chiu2006} can be derived 
as a special case of the extended bent-cable. } \\

\noindent Let $\psi(z) = (z+\gamma)^2/4\,\gamma$, then 
\begin{eqnarray}
\psi^{*}(x) = \frac{\psi(z')}{|\delta|} = \frac{[|\delta| (x-\tau)+\gamma]^2}{4\, \gamma |\delta|}. \nonumber
\end{eqnarray}
By dividing the numerator and denominator by $|\delta|^2$, we have that 
\begin{eqnarray}
\psi^{*}(x)  = \frac{(x-\tau+\gamma / |\delta|)^2}{4\, \gamma / |\delta|} = \frac{(x-\tau+\zeta)^2}{4\, \zeta}. \nonumber
\end{eqnarray}

\subsection{\label{proof1.4}The fourth-degree polynomial bent-cable}

{\it The adaptation of the Zang's suggestion for Tishler and Zang's approximation 
provides a $4^{th}$-degree polynomial bent-cable.}\\

\noindent The suggestion of Zang \cite{Zang1980} is given by
\begin{eqnarray}
\label{quartaordem}
\psi(z) = -\frac{z^4}{16 \, \gamma^3} + \frac{3\, z^2}{8 \, \gamma} + \frac{z}{2} + \frac{3 \, \gamma}{16} \nonumber. 
\end{eqnarray}
Thus, the corresponded (extended) bent-cable is given by 
\begin{eqnarray}
\psi^{*}(x) = \frac{\psi(z')}{|\delta|} = -\frac{(x-\tau)^4}{16 \, \zeta^3} + \frac{3\, (x-\tau)^2}{8 \, \zeta} + \frac{(x-\tau)}{2} + \frac{3 \, \zeta}{16}. \nonumber
\end{eqnarray}

\section{On the state mixture models}

\subsection{\label{proof2.1}Equivalence between sign and max-min formulations} 
{\it Replacing $u(x-\tau)$ with $(x-\tau) \, F_T(x)$ in max-min models is the same as replacing $\sgn(x-\tau)$ with $2 \, F_T(x) - 1$ in the sign formulation. Models obtained from both approaches are reparametrizations for the same curve.}\\

\noindent Let's consider the sign formulation. The proposed state mixture model is given by
\begin{eqnarray}
\eta(x)  = \theta_0 + \theta_1 (x-\tau) + \theta_2 (x-\tau) \, [ 2 \, F_{T}(x) - 1 ]. \nonumber
\end{eqnarray}
The model is rewritten as follows  
\begin{eqnarray}
\eta(x) &=& \alpha_l + \beta_l \, \tau + \frac{(x-\tau)}{2} \Big [ (\beta_l+\beta_{l+1}) +
(\beta_{l+1}-\beta_l) \, ( 2 \, F_{T}(x) - 1 ) \Big ] \nonumber \\
&=& \alpha_l + \beta_l \, x \, (1 - F_{T}(x)) + \left [\beta_{l+1} \, x + \tau \, (\beta_{l} - \beta_{l+1}) \right ] \, F_T(x). \nonumber
\end{eqnarray}
Since $\tau = (\alpha_l - \alpha_{l+1})/(\beta_{l+1} - \beta_{l})$, 
\begin{eqnarray}
\eta(x) = (\alpha_l+ \beta_l \, x) (1 - F_{T}(x) ) + (\alpha_{l+1}+ \beta_{l+1} \, x) \, F_{T}(x), \nonumber
\end{eqnarray}
resulting in a smoothed max-min model (Eq. \ref{mixture}). 

\subsection{\label{proof2.2} Theorem \ref{teo1} - Lemma 1.1}
{\it All non-decreasing, smooth transitional functions can be written as   
\begin{eqnarray}
\trn(x-\tau, \, \gamma)  =  2 F_T(x) - 1, ~~~ -\infty < \tau < +\infty, ~~ \gamma>0, \nonumber
\end{eqnarray}
where $T$ is a random threshold and $F_T(x)$ is its cumulative distribution function. 
} \\

\noindent The function $F_T(x)=[\trn(x-\tau, \, \gamma)+1]/2$ 
is a monotonous transformation of $\trn(x-\tau,\gamma)$, then it is non-decreasing. 
Also, the transitional function are smooth ($C^1$ class) and, in this way, 
$F_T(x)$ is a right-continuous function. By the last, the condition (i) for 
transitional models means that
\begin{eqnarray}
\lim_{x \rightarrow \pm \infty} (x-\tau)  [ \trn(x-\tau, \, \gamma) - \sgn(x-\tau) ]  = 0 \nonumber
\end{eqnarray}
which implies 
\begin{eqnarray}
\lim_{x \rightarrow \pm \infty} [ \trn(x-\tau, \, \gamma) - \sgn(x-\tau) ]  = 0 \nonumber
\end{eqnarray}
and, consequently, $F_T(x)$ is limited by 
\begin{eqnarray}
\displaystyle{\lim_{x \rightarrow -\infty} F_T(x) = 0}  ~~~~ \mbox{and} ~~~~ \displaystyle{\lim_{x \rightarrow +\infty} F_T(x) = 1}.  \nonumber     
\end{eqnarray}
Since $F_T(x)$ is right-continuous, non-decreasing, and limited in $[0,1]$, 
there is a random variable $T$ for which $F_T(x)$ is the 
cumulative distribution function. 

\subsection{\label{proof2.3} Theorem \ref{teo1} - Lemma 1.2}
{\it The state mixture model is equivalent to the use of a non-decreasing, smooth transitional function always when the random threshold $T$: (a) has a continuous PDF and (b) $E|T|<+\infty$ or its support is limited.} \\
 
\noindent The PDF is always non-negative and it ensures a non-decreasing $F_T(x)$. As a consequence, the resulting $\trn(.)$ is also non-decreasing. 
In order to obtain a smooth $\trn(.)$, $F_T(x)$ has to be continuously differentiable and, therefore, it requires a continuous PDF over $x$. In addition, the distribution must be 
such that the three conditions for transitional functions are satisfied. \\

\noindent Part 1: The validity of condition (i) for $\trn(.)$ is equivalent to 
\begin{eqnarray}
(c.1) \displaystyle{\lim_{x \rightarrow -\infty}} x \, F_{T}(x)  = 0 ~~~  \mbox{and} ~~~ (c.2) \displaystyle{\lim_{x \rightarrow +\infty}} x \, [F_{T}(x) - 1] = 0. \nonumber
\end{eqnarray}
This condition clearly holds for any distribution with proper and bounded support, since that for finite $x$ values the $F_T(x)$ function reaches its limits [$F_T(x_{inf}) = 0$ and $F_T(x_{sup}) = 1$]. However, for unbounded support, the limit indeterminacy must be solved. \\ 

\noindent At first, let's consider a fixed and negative $x$ when analysing $(c.1)$. In this case 
\begin{eqnarray}
\label{c1}
0 \, \geq \,  x \, F_T(x) = x \, \int_{-\infty}^{x} \, dF(\mu) \geq \int_{-\infty}^{x} \mu \, dF(\mu), \end{eqnarray}
since $x f_T(\mu) \geq \mu f_T(\mu)$ for any $\mu \in (-\infty, x]$. The right-side     
of the inequality can be expressed by 
\begin{eqnarray}
\int_{-\infty}^{x} \mu \, dF(\mu) = E[T] - \int_{x}^{+\infty} \mu \, dF(\mu)  \nonumber
\end{eqnarray}
and, if it has a limit when $x \rightarrow -\infty$, the limit will result 
in  
\begin{eqnarray}
\displaystyle{\lim_{x \rightarrow -\infty}} \Big ( E[T] - \int_{x}^{+\infty} \mu \, dF(\mu) \Big ) = 0. \nonumber
\end{eqnarray}
This limit holds only if $E[T]$ is well-defined, that is, when the random variable $T$ is integrable ($E|T|<+\infty$). Then, by taking the limits in both sides of Eq. (\ref{c1}), 
the validity of $(c.1)$ is proved using the squeeze theorem.\\

\noindent The limit $(c.2)$ is proved in a similar way. Let's consider a fixed and positive $x$.
Then, 
\begin{eqnarray}
\label{c2}
0 \, \leq \, - x \, [F_T(x) - 1]  ~ = ~ x \, \int_{x}^{+\infty} \, dF(\mu) \leq \int_{x}^{+\infty} \mu \, dF(\mu), 
\end{eqnarray}
since $x f_T(\mu) \leq \mu f_T(\mu)$ for any $\mu \in [x, +\infty)$. The right-side     
of the inequality is expressed by
\begin{eqnarray}
\int_{x}^{+\infty} \mu \, dF(\mu) = E[T] - \int_{-\infty}^{x} \mu \, dF(\mu).   \nonumber
\end{eqnarray}
Considering an integrable $T$, the limit bellow holds: 
\begin{eqnarray}
\displaystyle{\lim_{x \rightarrow +\infty}} \Big ( E[T] - \int_{-\infty}^{x} \mu \, dF(\mu) \Big ) = 0. \nonumber
\end{eqnarray}
Then $(c.2)$ is proved by taking the limit in both sides of the Eq. (\ref{c2}). \\ 

\noindent Part 2: The condition (ii) determines that $\trn(x-\tau, \gamma)$ behaves as $\sgn(x-\tau)$ for any $x \neq \tau$ when the parameter of curvature $\gamma$ approaches to zero. \\ 

\noindent Let's consider $\gamma$ (Eq. \ref{underdist}) proportional to the scale parameter, $\nu$. In this way, when $\gamma \rightarrow 0^{+}$ the PDF behaves like the Dirac's delta function  
\begin{eqnarray}
\Delta(x) = \left \{
\begin{array}{lc}
\infty & x = \tau, \\
0       & x \neq \tau,
\end{array}
\right . \nonumber
\end{eqnarray}
and the CDF is given by $\mbox{I}_{\{x \geq \tau \}} (x)$. Thus
\begin{eqnarray}
\label{teste}
\displaystyle{\lim_{\gamma \rightarrow 0^{+}}} 2 \, F_{T}(x) - 1 = 2 \, \mbox{I}_{\{x \geq \tau\}} (x) - 1 = \sgn(x-\tau)   ~~~ \forall \, x \neq \tau. \nonumber
\end{eqnarray}

\noindent Part 3: The condition (iii) holds since $\trn(0) = 2 \, F_T(0) - 1  \, \in \, [-1,1]$, that is, $\trn(0)$ has a finite value. \\

\noindent As a consequence of the {\it{Lemma 1.2}}, we have that 
\begin{eqnarray}
f_{T} (x | \tau, \, \gamma) = \frac{1}{2} \frac{d}{d x} \trn(x-\tau , \gamma) \nonumber
\end{eqnarray}
is the probability density function for the random threshold $T$ and 
\begin{eqnarray}
F_T(x) = P(T \leq x) = \int_{-\infty}^{x} f_{T} (x | \tau, \, \gamma) \, dx \nonumber
\end{eqnarray}
is the probability of phase-transition $l \rightarrow l+1$ has been occurred for a given $x$.

\subsection{\label{proof2.4} Logistic threshold and Bacon and Watts's model}
{\it The hyperbolic tangent as transitional function leads to the state mixture model with a Logistic random threshold $T$.} \\

\noindent Bacon and Watts \cite{Bacon1971} proposed the hyperbolic tangent as transitional function, as the following 
\begin{eqnarray}
\trn(x - \tau, \gamma) = \frac{e^{(x-\tau)/\gamma} + e^{-(x-\tau)/\gamma}}{e^{(x-\tau)/\gamma} - e^{-(x-\tau)/\gamma}}. \nonumber
\end{eqnarray}
It was noted that 
\begin{eqnarray}
F_T(x) = \frac{1}{2} \Big [ \trn(x - \tau, \gamma) + 1 \Big ] = 
\frac{1}{2} \Big \{ \frac{\sinh[(x-\tau)/\gamma] + \cosh[(x-\tau)/\gamma]}{\cosh[(x-\tau)/\gamma]} \Big \}, \nonumber
\end{eqnarray}
which results in the CDF for the Logistic random variable $T$, with $E(T)=\tau$ and 
dispersion parameter $\nu=\gamma/2$, that is given by
\begin{eqnarray}
F_T(x) = \frac{1}{\Big ( 1 + e^{-2\, (x-\tau) /\gamma} \Big )}. \nonumber
\end{eqnarray}

\subsection{\label{proof2.5} Threshold with Bi-weight distribution}
{\it The high degree polynomial suggested as transitional function by Bunke and Schulze and L\'azaro et al. \cite{Bunke1985,Lazaro2001} is compatible with a state mixture model based on the bi-weight distribution for $T$} \\

\noindent The bi-weight (or quartic) function is given by $K(t) = \frac{15}{16} (1-t^2)^2$. We can state a distribution based on this shape, and limited in $x \in [\tau-\zeta, \tau+\zeta]$, which is given by
\begin{eqnarray}
f_T(x) = \frac{15}{16 \, \zeta^5} [\zeta^2-(x-\tau)^2]^2 \, I_{[\tau-\zeta, \tau+\zeta](x)}. \nonumber
\end{eqnarray}
Consequently, the CDF is expressed by
{\small{
\begin{eqnarray}
F_T(x) = \Big ( \frac{1}{2} + \frac{15 \, (x-\tau)}{16 \, \zeta} - \frac{5\, (x-\tau)^3}{8 \, \zeta^3} + \frac{3 \, (x-\tau)^5}{16 \, \zeta^5} \Big ) I_{[\tau-\zeta,\tau+ \zeta]}(x) + I_{\{x>\tau+\zeta\}}(x). \nonumber
\end{eqnarray}
}}
Thus, by making $\psi^*(x) = (x-\tau) \, F_T(x)$ it is obtained the $6^{th}$-degree polynomial bent-cable derived from Lazaro's and Bunke and Schulze's approximations, 
which is given by
\begin{eqnarray}
\psi^{*}(x) = \frac{(x-\tau)}{2} + \frac{15 \, (x-\tau)^2}{16 \, \zeta} - \frac{5\, (x-\tau)^4}{8 \, \zeta^3} + \frac{3 \, (x-\tau)^6}{16 \, \zeta^5}.\nonumber
\end{eqnarray} 

\subsection{\label{proof2.6}Threshold with Cauchy's distribution} 
{\it There is no valid transitional approximation associated with a random threshold 
$T$ with Cauchy's distribution. }\\

\noindent Let's consider a random threshold $T$ distributed according to the Cauchy's distribution, with scale parameter $\nu = 1$. Then, the PDF is given by 
\begin{eqnarray}
F_T(x) = \frac{1}{\pi} \tan^{-1} (x-\tau) + \frac{1}{2}. \nonumber
\end{eqnarray}
Thus, the $\trn(x-\tau) = 2\, F_T(x) - 1$ violates the condition (i) since 
\begin{eqnarray}
\displaystyle{\lim_{x \rightarrow -\infty}}  x \, F_T(x) = \displaystyle{\lim_{x \rightarrow +\infty}}  x \, [F_T(x) - 1] = - \frac{1}{\pi}. \nonumber
\end{eqnarray}

\section{On the expected bending-cable}

\subsection{\label{proof3.1}Approximations for the abrupt operators} 
{\it For any integrable and continuous random variable $T$, the expected values for $u(x-T)$ and $|x-T|$ are given by Eq. (\ref{expectedmax}) and Eq. (\ref{expectedsignal}), respectively.} \\

\noindent Let's consider a translation for the random threshold $S=T-\tau$, centring its distribution. In this way, $F_T(x) = F_S(x-\tau)$ and $f_T(x) = f_S(x-\tau)$. Thus, 
{\small{
\begin{eqnarray}
E[u(x-T)] = E[u(x-S-\tau)] = E[\max\{x-S-\tau,0\}] = E[\max\{x-\tau, \, S\} - S]. \nonumber 
\end{eqnarray}
}}
For an integrable $T$, it results in
{\small{
\begin{eqnarray}
&& E[u(x-T)]  =  (x-\tau) \, \int_{-\infty}^{x-\tau} \, dF_S(S) + \int_{x-\tau}^{+\infty} S \, dF_S(S) -  \int_{-\infty}^{+\infty} S \, dF_S(S)  = \nonumber \\
&& (x-\tau) \, F_S(x-\tau) - \int_{-\infty}^{x-\tau} S \, dF_S(S)  =  (x-\tau) \, F_T(x) + \int_{x}^{+\infty} (T-\tau) \, dF_T(T). \nonumber
\end{eqnarray}
}}
In a similar way, $E|x-\tau|= E[(x-S-\tau) \, \sgn(x-S-\tau)]$ is given by
{\small{
\begin{eqnarray}
&&\int_{-\infty}^{x-\tau} (x - S - \tau) \, dF_S(S) - \int_{x-\tau}^{+\infty} (x-S-\tau) \, dF_S(S) = \nonumber \\
&& (x-\tau) \, \Big [ \int_{-\infty}^{x-\tau} \, dF_S(S) - \int_{x-\tau}^{+\infty} \, dF_S(S) \Big ] + \int_{x-\tau}^{\infty} S \, dF_S(S) - \int_{-\infty}^{x-\tau} S \, dF_S(S). \nonumber 
\end{eqnarray}
}}
Since 
\begin{eqnarray}
\int_{x-\tau}^{\infty} \, dF(S) = 1 - \int_{-\infty}^{x-\tau} \, dF(S), \nonumber
\end{eqnarray}
the expected value for the modulus is that 
\begin{eqnarray}
(x-\tau) \, \Big [ 2 \, F_T(x) - 1 \Big ] + \int_{x}^{\infty} (T-\tau) \, dF_T(T) - \int_{-\infty}^{x} (T-\tau) \, dF_T(T). \nonumber
\end{eqnarray}

\subsection{\label{proof3.2-2} On the asymptotic behaviour of the model}
{\it The term 
\begin{eqnarray}
\int_{x}^{+\infty} (T - \tau) \, dF_T(T) \nonumber
\end{eqnarray}
goes to zero at $x \rightarrow \pm \infty$ only when the random threshold $T$ is an integrable variable. }\\

\noindent For $x \rightarrow + \infty$ the term always goes to zero. But when $x \rightarrow - \infty$ we have that  
\begin{eqnarray}
\lim_{x \rightarrow - \infty} \int_{x}^{+\infty} (T - \tau) \, dF_T(T) = 
E(T - \tau) = E(T) -\tau = 0 \nonumber
\end{eqnarray}
only if $E(T)$ is well defined, i.e. only when $T$ is an integrable random variable. 

\subsection{\label{proof3.3} Theorem \ref{teo2}}
{\it The expected bending-cable models are equivalent to the use of hyperbolic approximation, satisfying the conditions (h.1 - h.4) and being expressed as
\begin{eqnarray}
\hyp(x-\tau, \gamma) =  2 \, F_T(x) - 1 - 2 \frac{\int_{-\infty}^{x} (T-\tau) \, dF_T(T)}{x-\tau} \nonumber.
\end{eqnarray}
Any integrable $T$ with a continuous PDF has an associated hyperbolic approximation given in this form. }\\

\noindent (h.1) {\it This proof was divided into two parts [conditions (i) and (ii)].} \\

\noindent Part 1 - the asymptotic behaviour of the any hyperbolic model is ensured by  
\begin{eqnarray}
(i) \lim_{x \rightarrow \pm \infty} \Big [ \hyp(x-\tau, \gamma) - |x-\tau| \Big ] = 0 \nonumber. 
\end{eqnarray}
In the case of the expected bending-cable models, the $\hyp(.)$ function is a 
transitional function with a correction factor, then it follows that the limit 
above is equal to 
\begin{eqnarray}
\lim_{x \rightarrow \pm \infty} \Big [\trn(x-\tau, \gamma) - 2 \, \frac{\int_{-\infty}^x (T-\tau) dF_T(T)}{x-\tau}  - |x-\tau| \Big ] = 0 \nonumber.
\end{eqnarray}
We know that transitional functions satisfy the condition (i). Thus, in order to condition (i) holds also for $\hyp(.)$, the correction factor must vanish when 
$x$ goes to infinity, that is: 
\begin{eqnarray}
\lim_{x \rightarrow \pm \infty} - 2 \, \frac{\int_{-\infty}^x (T-\tau) dF_T(T)}{x-\tau} = 0. \nonumber
\end{eqnarray}
The condition is met for any integrable  $T$ (see section \ref{proof3.2-2}).  \\

\noindent Part 2 - When $\gamma \rightarrow 0^{+}$ the PDF for $T$ behaves like the Dirac's delta function (see section \ref{proof2.3} - part 2). In this case the CDF is given by $\mbox{I}_{\{x \geq \tau \}} (x)$ and 
\begin{eqnarray}
\displaystyle{\lim_{\gamma \rightarrow 0^{+}}} \trn(x-\tau, \gamma) = 2 \, \mbox{I}_{\{x \geq \tau\}} (x) - 1 = \sgn(x-\tau)   ~~~ \forall \, x \neq \tau. \nonumber
\end{eqnarray}
Also, the correction term vanishes since 
\begin{eqnarray}
\lim_{\gamma \rightarrow 0^{+}} \int_{-\infty}^x (T-\tau) \, dF_T(T)
=  \int_{-\infty}^x (T-\tau) \, \Delta(T)  \, dT = 0 ~~~ x\neq\tau, \nonumber  
\end{eqnarray}
where $\Delta(.)$ is the Dirac's function centred at $\tau$. In this way, 
\begin{eqnarray}
\lim_{\gamma \rightarrow 0^{+}} 2 \, F_T(x) - 1 - 2 \frac{\int_{-\infty}^{x} (T-\tau) \, dF_T(T)}{x-\tau} =  \sgn(x-\tau)   ~~~ \forall \, x \neq \tau. \nonumber
\end{eqnarray}
In other words, the condition (ii) holds.   \\

\noindent (h.2) {\it For hyperbolic models the $\hyp(x-\tau = 0)$ has no finite value.} \\

\noindent It is easy to see that 
\begin{eqnarray}
\hyp(0) = \trn(0) - 2 \, \frac{\int_{-\infty}^\tau (T-\tau) dF_T(T)}{\lim_{x \rightarrow \tau} (x-\tau)}. \nonumber 
\end{eqnarray}
Since the limit in denominator goes to zero, $\hyp(0)$ is not defined and its limits diverges to infinity. \\

\noindent (h.3) {\it The third condition establishes that \it $|\hyp(x-\tau, \gamma)|\geq1$.}\\

\noindent For the expected bending-cable models, we have that 
\begin{eqnarray}
(x - \tau) \, \hyp(x-\tau, \gamma) = (x-\tau) \Big [ 2\, F_T(x) - 1 \Big] - 2 \, 
\int_{-\infty}^x (T-\tau) dF_T(T). \nonumber
\end{eqnarray}
It was verified that
\begin{eqnarray}
2\, (x-\tau) \, F_T(x) = 2 \, (x-\tau) \, \int_{-\infty}^x \, dF_T(T) \geq  2 \, 
\int_{-\infty}^x (T-\tau) \, dF_T(T), \nonumber
\end{eqnarray}
since $(T-\tau) \leq (x-\tau)$ for all the $T \leq x$. The $\hyp(.)$ function
was rewritten based on the difference of the two integrals above, considering 
the following quantity 
\begin{eqnarray}
d(x) = 2 \Big [ (x-\tau) \, F_T(x) -  
\int_{-\infty}^x (T-\tau) \, dF_T(T) \Big ] \geq 0, \nonumber
\end{eqnarray}
in such the manner that
\begin{eqnarray}
\label{hyplim}
\hyp(x-\tau) = \frac{d(x)}{x-\tau} - 1 
\end{eqnarray}
We analyse the behaviour of $d(x)/(x-\tau)$ in order to define limits 
for $\hyp(.)$. At first, it was verified that 
\begin{eqnarray}
\lim_{x \rightarrow \tau^{-}} \frac{d(x)}{x-\tau} = - \infty ~~~
\mbox{and}  ~~~ \lim_{x \rightarrow \tau^{+}} \frac{d(x)}{x-\tau} = + \infty, \nonumber
\end{eqnarray}
since $\int_{-\infty}^x (T-\tau) \, dF_T(T) \leq 0$ for any integrable $T$. 
Also, the first derivative of $d(x)/(x-\tau)$ is negative for all the 
$x\neq\tau$ and thus this function is monotonically decreasing for 
$x \in (-\infty, \tau)$ or $x \in (\tau,+\infty)$. Still considering an integrable $T$, 
the horizontal asymptotes for this function are given by
\begin{eqnarray}
\lim_{x \rightarrow -\infty} \frac{d(x)}{x-\tau} = \lim_{x \rightarrow -\infty} 
2 \int_{-\infty}^x \, dF_T(T)  = 0 \nonumber \\
\lim_{x \rightarrow +\infty} \frac{d(x)}{x-\tau} = \lim_{x \rightarrow +\infty} 
2 \int_{-\infty}^x \, dF_T(T)  = 2. \nonumber
\end{eqnarray}
In this way, the image set is $Im = (0,-\infty)$ for $x<\tau$, and 
$Im = (2,+\infty)$ for $x>\tau$. Then, by Eq. (\ref{hyplim}) 
follows that $|\hyp(x-\tau, \gamma)| \geq 1$.  \\

\noindent (h.4) {\it The fourth condition is satisfied when  $(x-\tau) \, \hyp(x-\tau, \gamma)$ is differentiable over $x \in \R$.} \\

\noindent By differentiating $(x-\tau) \, \hyp(x-\tau, \gamma)$ with respect to $x$, it is obtained
\begin{eqnarray}
\frac{d}{dx} \Big [(x-\tau) ( 2\, F_T(x) - 1 ) - 2 \, 
\int_{-\infty}^x (T-\tau) dF_T(T) \Big ] = 2\, F_T(x) - 1. \nonumber
\end{eqnarray}
Then, the fourth condition holds for any continuous $T$. 

\subsection{\label{prooffinal1}The hyperbolic model and the t-distribution}
{\it The Griffiths and Miller's hyperbolic model \cite{Griffiths1973} is a special case of the expected bending-cable models when $T$ has a non-standard t-distribution with 2 degrees of freedom.}\\

\noindent From the hyperbolic approximation, it is obtained that
\begin{eqnarray}
F_T(x) = \frac{(\sqrt{(x-\tau)^2+\gamma})^{\prime}+1}{2} = \frac{1}{2} \Big ( \frac{x-\tau}{\sqrt{(x-\tau)^2+\gamma}}  + 1\Big ), \nonumber
\end{eqnarray}
and, consequently, it follows that
\begin{eqnarray}
f_T(x) = \frac{\gamma}{2} \Big [ (x-\tau)^2+\gamma \Big ]^{-3/2}. \nonumber
\end{eqnarray}
This expression corresponds to the PDF for the non-standard t-distribution 
\begin{eqnarray}
f_T(x|\nu, \tau, \sigma) = \frac{\Gamma(\frac{\nu+1}{2})}{\Gamma(\nu/2) \, \sqrt{\nu \, \pi} \, \sigma } \Big [ 1 + \frac{1}{\nu} \Big ( \frac{x-\tau}{\sigma} \Big )^2 \Big ]^{- \frac{\nu + 1}{2}}, \nonumber
\end{eqnarray}
with $\nu=2$ and $\gamma = 2\, \sigma^2$. 

\subsection{\label{prooffinal2} The log and exponential model and the Logistic threshold} 
{\it The log and exponential $\lne(.)$ approximation \cite{Jimenez-Fernandez2016a} is 
a special case of the expected bending-cable when $T$ has a Logistic distribution.
}\\

\noindent As it was previously seen, $\lne(x-\tau, \gamma) = (x-\tau) \, \hyp(x-\tau, \gamma)$ and, in this 
way, 
\begin{eqnarray}
F_T(x) = \frac{\lne^{\prime}(x-\tau, \gamma)+1}{2} = \frac{1}{1+e^{-(x-\tau)/\gamma}}. \nonumber
\end{eqnarray}
Then, $F_T(x)$ is the CDF for Logistic distribution with location and scale determined 
by $\tau$ and $\gamma$, respectively.

\subsection{\label{outra}Bent-cable and Uniform threshold}
{\it The bent-cable model is a special case of the expected bending-cable model when 
$T$ has an Uniform distribution.}  \\

\noindent Let's consider $T \sim U(\tau - \zeta, \tau +\zeta)$. Thus, 
\begin{eqnarray}
f_T(x) = \frac{1}{2\, \zeta} I_{\{|x-\tau| \leq \zeta\}} (x), \nonumber 
\end{eqnarray}
and
\begin{eqnarray}
F_T(x) = 
\frac{(x-\tau+\zeta)}{4\, \zeta} \, I_{\{|x-\tau| \leq \zeta\}} (x) + I_{\{x> \tau + \zeta\}} (x). \nonumber
\end{eqnarray}
Then, we have that $E[u(x-T)]$ is equal to 
\begin{eqnarray}
\Big [ (x-\tau) \, \frac{(x-\tau+\zeta)}{2 \, \zeta}  + \int_{x}^{\tau+\zeta} \frac{(t-\tau)}{2 \, \zeta} dt  \Big ] \, I_{\{|x-\tau| \leq \zeta\}} (x) + (x-\tau) \, I_{\{x> \tau + \zeta\}} \nonumber . 
\end{eqnarray}
But 
\begin{eqnarray}
\int_{x}^{\tau+\zeta} \frac{(t-\tau)}{2 \, \zeta} dt = \frac{\zeta^2 - (x - \tau)^2}{4 \, \zeta}, \nonumber
\end{eqnarray}
and 
\begin{eqnarray}
(x-\tau) \, \frac{(x-\tau+\zeta)}{2 \, \zeta}  +  \frac{\zeta^2 - (x - \tau)^2}{4 \, \zeta} 
= \frac{(x-\tau)^2 + 2\,(x-\tau) \, \zeta + \zeta^2}{4 \, \zeta}. \nonumber
\end{eqnarray}
In this way, we have that
\begin{eqnarray}
E[u(x-T)] = \frac{(x-\tau+\zeta)^2}{4\, \zeta}  \, I_{\{|x-\tau| \leq \zeta\}} (x) + (x-\tau) \, I_{\{x> \tau + \zeta\}} \nonumber . 
\end{eqnarray}
Replacing it in the Eq. (\ref{expectedbent}), we obtain the original bent-cable model.\\ 

\noindent Alternatively, it can proved by the equation (\ref{verify}):
\begin{eqnarray}
\frac{d}{dx} \psi^{*}(x) = \frac{d}{dx}  \frac{(x-\tau + \zeta)^2}{4 \, \zeta}  = \frac{x - (\tau - \zeta)}{2 \, \zeta}. \nonumber
\end{eqnarray}
Note that the right-side of the equation above corresponds to the CDF for $T \sim U(\tau-\zeta, \tau + \zeta)$. 

\subsection{\label{proof3.5} The threshold with Epanechnikov's distribution}

{\it The $4^{th}$ degree polynomial bent-cable (Eq. \ref{ex-quartaordem}) is equal to the expected bending-cable model based on the Epanechnikov's distribution for $T$. }\\

\noindent Let's consider the Epanechnikov's distribution defined 
in $x \in [\tau-\zeta, \tau +\zeta]$. In this case
\begin{eqnarray}
f_T(x) = \frac{3 \, [\zeta^2 - (x-\tau)^2]}{4 \, \zeta^3} ~~~~\mbox{and} ~~~~ F_T(x) = - \frac{(x-\tau)^3}{4\,\zeta^3} + \frac{3 \, (x-\tau)}{4\, \zeta} + \frac{1}{2}, \nonumber 
\end{eqnarray}
for $\tau - \zeta < x < \tau + \zeta$. 
Note that the bent-cable in Eq. (\ref{quartaordem}) is uniquely related to this Epanechnikov's distribution, since
\begin{eqnarray}
\frac{d}{dx} \psi^{*}(x) = \frac{d}{dx} \Big ( -\frac{(x-\tau)^4}{16 \, \zeta^3} + \frac{3\, (x-\tau)^2}{8 \, \zeta} +
\frac{(x-\tau)}{2} + \frac{3 \, \zeta}{16} \Big ) =  F_T(x),  \nonumber
\end{eqnarray}
for $\tau - \zeta < x < \tau + \zeta.$.

\subsection{\label{proofunif} The Khan and Kar's model and the Exponentiated Uniform distribution} 
{\it The Khan and Kar's \cite{Khan2018} generalised bent-cable model is a particular case of the 
expected bending-cable model when $T$ has the Exponentiated Uniform distribution. }\\

\noindent For the generalised bent-cable model, the bent function is given by 
\begin{eqnarray}
\psi^*(x) = \frac{\zeta \, (x - \tau + \zeta \, (k-1))^k}{(\zeta \, k)^k}, ~~~  \tau - (k-1)\, \zeta<x\leq \tau + \zeta,  \nonumber
\end{eqnarray}
where $z_1 = \tau - (k-1) \, \zeta$, $z_2 = \tau + \zeta$, and $\tau = [z_2 \, (k-1)+z_1]/k$, with $k>1$. The bent can be rewritten as follows
\begin{eqnarray}
\psi^*(x) =  \frac{(x-z_1)^k}{k\,(z_2-z_1)^{k-1}}, ~~~  z_1 <x\leq z_2.   \nonumber
\end{eqnarray}
In this way, we have that
\begin{eqnarray}
\frac{d}{dx} \psi^*(x) =  \Big ( \frac{x-z_1}{z_2-z_1} \Big )^{k-1} = F_T(x), ~~~  z_1 <x\leq z_2.   \nonumber
\end{eqnarray}
In this case $F_T(x)$ is the CDF for the Exponentiated Uniform distribution according to 
Ramires et al. \cite{Ramires2019}, with $\alpha=k-1$.   
\end{document}